# Valorização da ciência e do ensino de Física através da comunicação teatral

## Valuing science and the teaching of Physics through theatrical communication

Ederaldo Bueno de Macedo Junior[1], Edemar Benedetti Filho[2] e James Alves de Souza[2,*]

[1] Escola Estadual Coronel Venâncio, Mogi Mirim, São Paulo, Brasil
[2] Universidade Federal de São Carlos, Sorocaba, São Paulo, Brasil
* jasouza@ufscar.br

**Resumo**
A comunicação científica dentro e fora das salas de aula é o principal meio para fornecer uma compreensão adequada sobre como a ciência e a inovação tecnológica se relacionam com a sociedade. Para que essa comunicação seja efetiva é necessário explorar novos instrumentos didáticos para o ensino de ciências de maneira geral. A intersecção entre ciências e artes tem contribuído para tornar as ciências acessíveis, relevantes e interessantes para públicos diversos, com potencial de fornecer conteúdo científico, valores artísticos e de entretenimento significativos. Neste trabalho utilizamos a comunicação teatral para valorizar a ciência e o ensino de Física através de um roteiro autoral de uma peça em que são abordados conceitos e fenômenos relacionados à astronomia. A partir de uma linguagem artística e descontraída adaptamos os conceitos científicos propostos ao caráter contextual e instrumental do teatro para mostrar o seu valor educativo e a sua capacidade de transformar a ciência em entretenimento. Nossa proposta envolveu os estudantes em um ambiente altamente motivador e colaborativo, permitindo que eles aprendessem a reconhecer, analisar e imaginar explicações e performances para comunicar a ciência para o público geral de forma criativa e alternativa.
**Palavras-chave:** teatro científico; astronomia; divulgação científica; ensino de física.

**Abstract**
Scientific communication inside and outside the classroom is the main means for providing an adequate understanding of how science and technological innovation relate to society. In order to achieve this goal, it is important to explore new didactic instruments for teaching science in general. The intersection between science and arts has contributed to make science accessible, relevant and interesting to diverse audiences with the potential to provide scientific content, significant artistic and entertainment values. Here, we use theatrical communication to value science and the teaching of Physics through an original script of a play in which concepts and phenomena related to astronomy are addressed. Using an artistic and casual language, we adapt the proposed scientific concepts to the contextual and instrumental nature of theater to demonstrate its educational value and its potential to transform science into entertainment. Our proposal allowed students being engaged in a highly motivating and collaborative environment, providing them the opportunity to recognize, analyze and imagine explanations and performances to communicate science to the general public in a creative and alternative way.

# 1. Introdução

A ciência é uma atividade profissional coletiva e transdisciplinar desenvolvida por cientistas movidos por uma curiosidade profunda e pelo desejo de compreender os fenômenos naturais através de hipóteses e teorias que podem ser testadas e refutadas. Apesar da ciência ser um campo dinâmico que busca revelar as leis da natureza de maneira objetiva, ela é influenciada por contextos culturais, sociais, políticos e econômicos, que podem impactar a forma como as questões científicas são formuladas e investigadas. A tecnologia, por sua vez, além de focar na aplicação prática do conhecimento científico para resolver problemas específicos ou transformar a realidade de maneira concreta, é também uma ferramenta que facilita e impulsiona a pesquisa científica [1, 2, 3]. A ciência pode ser aplicada de maneira positiva para melhorar as nossas vidas através da medicina, na cura e prevenção de doenças, no desenvolvimento de novas vacinas, na eficiência de transportes, na geração de novas fontes de energia, na proteção do meio ambiente, na produção de alimentos, na transmissão de informação, na melhoria da educação, entre outros. A aplicação da ciência, através da tecnologia, também possui efeitos indesejados para a sociedade, pois contribui significativamente para a poluição do meio ambiente com as atividades industriais, que emitem gases e resíduos para a atmosfera e rios destruindo ecossistemas, gerando desperdício e consumo de recursos naturais. Outro ponto negativo é a produção de armas que destrói vidas e oprime nações menos favorecidas.

Para envolver as próximas gerações de estudantes na aprendizagem de ciências e na aplicação consciente e responsável da mesma, através do desenvolvimento de novas tecnologias, é imprescindível explorar novas metodologias de ensino para introduzir os conceitos científicos de uma maneira que a ciência seja reconhecida como algo relevante em nossas vidas, elencando a forma como estamos vivendo. Neste sentido, o teatro tem um potencial significativo para explorar o papel da ciência e da tecnologia na vida cotidiana, podendo fornecer meios para aprofundar as percepções dos estudantes sobre a condição humana e criar novas formas de aprendizado tanto para as ciências quanto para as artes.

O teatro como instrumento educativo no Brasil vem ganhando espaço desde que a Lei nº. 13.278 foi decretada em 2016 para incluir o mesmo, juntamente com as artes visuais, a dança e a música, como linguagem no componente curricular dos diversos níveis da educação básica [4]. Este se caracteriza como um importante meio de comunicação e expressão que articula aspectos audiovisuais, musicais e linguísticos com capacidade de influenciar e promover diferentes abordagens de temas educacionais nas variadas áreas do conhecimento, para a melhoria do desenvolvimento cognitivo e afetivo dos estudantes [5, 6, 7].

O teatro com temática científica, expressão cunhada por Moreira e Marandino [8], fornece o cenário ideal para promover a sinergia entre ciência, ética e sociedade. Nas últimas décadas vários trabalhos foram publicados para mostrar os benefícios e potencialidades da implementação do teatro em espaços formais e não formais para o ensino de ciências. Dentre estes, podemos destacar a utilização do mesmo para divulgação e popularização da ciência [9, 10], para a implementação da alfabetização científica [8], para a melhoria da formação docente [11, 12], para análise de conteúdo e concepções alternativas dos estudantes [13, 14, 15] e como estratégia didática para introdução e desenvolvimento de conceitos científicos e aspectos históricos da ciência [16, 17, 18, 19, 20, 21].

Grande parte dos trabalhos citados utilizaram os jogos teatrais como metodologia de ensino, cuja base pedagógica é a improvisação. A oportunidade de improvisação oferecida pelo teatro pode ser benéfica para ajustar, adaptar ou revelar para o professor conceitos que estão corretos ou parcialmente corretos na mente do estudante em relação à perspectiva formal fornecida em sala de aula. Se estes conceitos forem trabalhados de maneira adequada, em vez de serem simplesmente rejeitados ou rotulados como incorretos, a aprendizagem científica pode ser mais efetiva [22, 23, 24].

Neste trabalho apresentamos e discutimos o roteiro autoral de uma peça intitulada como "Sistema Maluco" para a representação teatral de fenômenos e conceitos científicos relacionados a área da astronomia. Além da transposição didática e adaptação dos conceitos propostos ao caráter contextual e instrumental do teatro, nós mostramos como é possível valorizar a ciência e o ensino de Física através de uma linguagem artística e mais descontraída. Adicionalmente, compartilhamos a experiência que adquirimos com duas apresentações feitas para o público geral em uma feira de ciências de uma escola particular do interior do estado de São Paulo. Estas contribuíram para que o roteiro autoral apresentado no material complementar deste trabalho e toda a sua trama fossem otimizados para que nossa proposta possa ser aplicada tanto no contexto de sala de aula quanto para eventos que promovam a disseminação e a popularização da ciência.

## 2. Escolha do tema e preparação da peça teatral: "Sistema Maluco"

O nome "Sistema Maluco" escolhido para a peça de teatro faz alusão ao nosso sistema solar. A palavra "maluco" é uma gíria muito utilizada entre os adolescentes brasileiros como brincadeira para se referirem a algo complexo ou cujo entendimento não seja direto. A Física recebe usualmente essa conotação pelos estudantes do ensino médio devido à dificuldade de entendimento, falta de interesse e baixo nível de aproveitamento que os mesmos apresentam com relação a esta disciplina [25].

Apesar das dificuldades aparentes, elencadas pelos próprios estudantes, tentamos transmitir a mensagem de que a ciência também pode ser abordada de maneira divertida e descontraída em uma peça de teatro.

Como o nosso objetivo inicial foi atingir o público geral, a escolha da astronomia como tema norteador do roteiro da peça surgiu naturalmente, pois esta é uma área que permite mostrar amplamente que a Física é uma ciência cotidiana. Adicionalmente, a astronomia está bem posicionada nos meios de comunicação social, com capacidade de despertar o interesse de um grande número de pessoas pela sua natureza excitante através de uma variedade de eventos de divulgação pública em todo o mundo [26].

Na peça "Sistema Maluco" os corpos celestes, como a Lua, o Sol e o planeta Terra, são os protagonistas. Eles dialogam entre si para explicar como o sistema solar funciona e porque observamos alguns dos fenômenos mais comuns do nosso cotidiano, como as estações do ano, os dias e as noites, as fases da Lua, os eclipses, entre outros. Toda a exposição de conceitos foi feita de uma maneira descritiva, para tornar os mesmos mais acessíveis e também para despertar a curiosidade de todos pela ciência.

Uma pergunta que pode surgir naturalmente para o leitor é: por que realizar um teatro considerando corpos celestes como os principais personagens e não pessoas, como cientistas que revolucionaram a nossa forma de pensar sobre o universo e o seu funcionamento?

Nosso propósito foi apresentar a ciência em sua própria definição, como um engajamento ativo e criativo de nossas mentes com a natureza, com o objetivo de compreendê-la de forma provisória e revisável. Esse engajamento se dá formalmente através de métodos como a observação, a experimentação e as análises matemáticas, mas também envolve interpretações teóricas e revisões, que são influenciadas constantemente por contextos sociais, culturais e históricos. O teatro oferece a oportunidade para qualquer pessoa, mesmo aquelas sem nenhuma formação básica ou técnica em ciências, de explorar o método da observação de uma maneira lúdica e divertida. Nessa abordagem, a "natureza" estabelece um diálogo indireto com a audiência para revelar os seus mistérios e para mostrar algumas das consequências decorrentes do nosso comportamento e de nossas atitudes com relação ao meio ambiente. Além da valorização da ciência, a rejeição ao negacionismo científico também é considerada em toda a trama.

No material suplementar deste artigo apresentamos um roteiro autoral completo da peça que pode ser aplicado tanto em um contexto de sala de aula quanto para eventos escolares, como feiras de ciências. A peça foi dividida em 8 esquetes para facilitar os ensaios e a assimilação dos conceitos pelos estudantes-atores. Um esquete é uma cena de curta duração com poucos atores. Esta estratégia também foi útil para não sobrecarregar a audiência com um excesso de informações, permitindo que o conhecimento fosse transmitido e organizado de maneira objetiva e mais simples.

O roteiro da peça foi preparado para permitir a atuação de até 40 estudantes, se cada esquete for apresentado por um grupo de estudantes diferente. Isso viabiliza a consideração da peça em um contexto de sala de aula pelo professor, para que todos os estudantes de uma turma possam participar.

A aplicação de nossa proposta foi feita em uma feira de ciências de uma escola particular do interior do estado de São Paulo com duas apresentações para o público geral. A escolha do elenco foi feita através de um convite feito para todos os estudantes de duas turmas do 1º ano do ensino médio da escola. Não houve nenhum processo seletivo ou qualquer obrigatoriedade para forçar a participação dos mesmos. Contudo, apenas 15 estudantes se prontificaram a participar. Uma vez que outras atividades científicas estavam sendo preparadas para a feira, os estudantes da escola, de maneira geral, tiveram a liberdade de escolher o que fariam.

Os ensaios foram realizados no período contrário das aulas na própria escola. No primeiro encontro foi feito o planejamento da atividade para a organização da peça, a exposição do tema e a forma como os estudantes poderiam contribuir e participar da mesma. No segundo encontro iniciamos os ensaios, considerando os temas e os conceitos que seriam abordados nos primeiros esquetes e a escolha das personagens. Diante da necessidade de dedicação aos ensaios e a condução de estudos para o entendimento e discussão dos conceitos físicos, vários estudantes desistiram permanecendo apenas 5 para apresentar todo o roteiro da peça. Com este número reduzido foi necessário simplificar consideravelmente as falas das personagens, para que elas conseguissem apresentar todos os conceitos propostos [27].

Vale ressaltar que durante as apresentações realizadas os estudantes memorizaram todas as suas falas independentemente. Com o roteiro atualizado, aumentamos o número de falas das personagens para fornecer melhores contextos e explicações para a audiência. A ansiedade e o medo do palco aliados a um grande número de falas, ou falas longas, podem fazer com que os estudantes-atores esqueçam falas, cometam erros ou percam o ritmo da apresentação. Para proporcionar maior conforto, confiança e facilitar a atuação dos estudantes durante as apresentações, o professor pode avaliar a possibilidade de conduzir a peça com o uso de cartões que contenham palavras-chave, escolhidas pelos próprios estudantes.

Para fazer com que os estudantes se sentissem mais motivados a aprenderem os conceitos de Física, o professor conduziu os ensaios de maneira descontraída, com brincadeiras, sempre estimulando os mesmos a fazerem improvisações em suas falas e atuações. Um dos propósitos de nossa peça é apresentar a ciência ao público de maneira divertida. Essa abordagem favoreceu o estabelecimento de uma dinâmica mais independente entre os estudantes. Eles privilegiaram o entendimento e a exposição dos conceitos científicos com suas próprias palavras, evitando a memorização do texto original. Isso foi intensamente valorizado pelo professor, que teve o trabalho adicional de analisar a consistência das falas improvisadas para que não houvessem erros na abordagem dos conceitos propostos. A improvisação fez com que os estudantes se envolvessem intelectualmente, fisicamente e intuitivamente com o tema da peça e isso foi essencial para a realização dos movimentos corretos nas cenas, pois para isso é imprescindível que as personagens entendam os conceitos [23]. Dessa forma, o professor conseguiu promover o teatro como uma possibilidade de estudo mais aprofundada e contextualizada dos conceitos científicos.

Na próxima seção explicamos e justificamos como cada esquete da peça foi pensado para comunicar ao público o entusiasmo e os benefícios da evolução do conhecimento através da ciência e da tecnologia com o tema astronomia, elencando as falas, as expressões e os movimentos dos estudantes-atores. Estes foram otimizados e ampliados a partir da experiência dos dois espetáculos conduzidos na escola [27].

## 3. Luz, Física e Ação

Em cada esquete a seguir apresentamos uma explicação detalhada dos conceitos físicos abordados na peça, com o objetivo de complementar aqueles considerados no roteiro para fornecer um melhor entendimento sobre a performance dos estudantes-atores e apresentar sugestões para a criação de novos esquetes e possibilidades de atuação.

### *Esquete 1: Apresentando o "Sistema Maluco" e o Modelo Geocêntrico*

O espetáculo foi iniciado com uma introdução feita pelo narrador para contextualizar o nome da peça "Sistema Maluco" e, principalmente, para apresentar a ciência como uma atividade que vem sendo aprimorada desde a antiguidade.

Adicionalmente, foi apresentado o que é Astronomia para explicar para o público a área da ciência que foi trabalhada na peça e para diferenciar a atividade de observação do céu feita por qualquer pessoa e por um astrônomo, que é o cientista que atua na área. Tivemos o cuidado de colocar a ciência em foco sempre que possível para reforçar a ideia de que a mesma não está relacionada a crenças ou conveniências de grupos de pessoas. Apesar da ciência não estar isenta de influências sociais, culturais e históricas, esta é uma atividade desenvolvida por profissionais capacitados e altamente treinados.

Em seguida o narrador introduz o primeiro modelo de sistema solar, conhecido como teoria do universo geocêntrico ou geocentrismo, proposto no século II da era cristã pelo astrônomo e geógrafo grego Cláudio Ptolomeu [28].

No início de todos os esquetes o narrador faz uma pequena introdução sobre o tema a ser trabalhado pelos estudantes-atores para situar a audiência cronologicamente sobre alguns acontecimentos, apresentar curiosidades, os problemas e desafios da época, assim como perguntas provocativas visando despertar e manter o interesse da audiência pela peça.

Os principais personagens da peça são a Terra, a Lua e o Sol. Os outros planetas do sistema solar foram representados por um único estudante-ator e foram considerados apenas nos 3 primeiros esquetes para apresentar a diferença entre o modelo geocêntrico de Ptolomeu e o modelo heliocêntrico de Copérnico. O referencial inercial do "Sistema Maluco", descrito pelas estrelas do firmamento, também é mencionado e representado no palco por uma grande cortina de helanquinha preta repleta de estrelas coloridas e feitas de cartolina.

O narrador chama a atenção da audiência para a esfericidade da Terra, neste primeiro momento, ao estabelecer a região de observação do céu noturno como sendo o Hemisfério Sul de nosso planeta. Isso é importante, porque as estrelas no firmamento não são as mesmas para todas as regiões do planeta. Isso significa que, uma pessoa que mora no Brasil, no Hemisfério Sul, e outra que mora nos Estados Unidos, no Hemisfério Norte, por exemplo, verão, simultaneamente, conjuntos de estrelas diferentes no firmamento.

Essa é uma evidência interessante para contestar o movimento terraplanista, em que as pessoas defendem que a Terra é plana com argumentos não científicos e teorias de conspiração [29]. Este movimento ganhou força a partir de 2014 através de grandes grupos de pessoas e canais de divulgação de conteúdos não científicos na internet.

Se a Terra fosse plana, todos teriam condições de observar o mesmo céu noturno simultaneamente, mesmo estando em diferentes regiões do planeta. Se o professor desejar, o narrador pode discursar sobre outras evidências da esfericidade da Terra no primeiro esquete, que podem ser observadas por qualquer pessoa a olho nu, como a existência de horizonte, o desaparecimento progressivo de navios no horizonte, o amanhecer e o anoitecer, a diferença da duração dos dias e das noites em diferentes lugares do planeta, a existência de fuso-horários e as diferenças das estações do ano ao longo do planeta. Isso pode ser feito de maneira introdutória, uma vez que vários desses assuntos são tratados em outros esquetes. O narrador pode ainda mencionar outras formas de verificar a esfericidade da Terra através de tecnologias atuais utilizadas em missões espaciais e balões atmosféricos que permitem tirar fotos do planeta, o uso de sistemas de navegação, planos de voo, entre outros.

A encenação é iniciada com o estudante-ator que interpreta a Terra entrando e se posicionando no centro do palco, para representar que a mesma ocupa o centro do universo. Posteriormente, a Lua, o Sol e os Planetas entram em cena, sucessivamente, realizando um movimento circular no sentido horário em torno da Terra. Eles se apresentam, dizem que giram em torno de nosso planeta, que se mantém imóvel, através de órbitas circulares. Apesar do geocentrismo ser um modelo para descrever o sistema solar, para as pessoas que defendiam a teoria geocêntrica na antiguidade, a configuração apresentada pelos estudantes-atores representava o próprio universo.

Todos os esquetes foram finalizados com os estudantes-atores saindo do palco para que a atenção fosse voltada exclusivamente para o narrador, durante a introdução do esquete seguinte.

O modelo geocêntrico se manteve até o final do século XVI e é muito utilizado ainda hoje pelos terraplanistas. Mas por que um modelo, que sabemos definitivamente que não reproduz a realidade do movimento relativo dos planetas, estrelas e outros corpos celestes, se manteve por tantos séculos?

Isso ocorreu e ainda ocorre com os terraplanistas porque o geocentrismo reproduz de maneira satisfatória aquilo que vemos. Se considerarmos a observação do céu a olho nu, qualquer pessoa pode ser levada a acreditar nesse modelo, pois o que percebemos é o planeta Terra, ou seja, nós mesmos, no centro de tudo que conseguimos ver e em repouso. Observamos a Lua e o Sol se movendo no céu e uma infinidade de pontos brilhantes fixos no firmamento.

Diante desse cenário, como é possível convencer o público geral que, aquilo que vemos não é necessariamente a realidade da natureza?

Para isso, é importante entender e reconhecer a ciência como uma atividade investigativa contínua, que se mantém em constante evolução, com momentos de estabilidade e rupturas paradigmáticas que desafiam e aprimoram a compreensão dos fenômenos que observamos. Isto significa que, além da ciência aprimorar seus métodos, modelos e teorias, estes são constantemente desafiados e revisados à medida que novas evidências e abordagens surgem.

Os astrônomos da antiguidade, que eram os cientistas da época, observavam outros corpos celestes além do Sol e da Lua. Eles sabiam da existência de 5 planetas: Mercúrio, Vênus, Marte, Júpiter e Saturno. Ao longo de 1 ano, por exemplo, eles observavam que o movimento do Sol e da Lua parecia simples e ordenado e podia ser previsto com o uso do modelo geocêntrico de maneira consistente com a realidade. Contudo, o movimento dos 5 planetas neste mesmo período de tempo não era tão fácil de entender, pois não podia ser explicado ou mesmo previsto pelo geocentrismo. Os planetas variavam o seu brilho, não mantinham uma posição fixa no céu, pareciam acelerar e desacelerar durante o movimento e, em algumas vezes, eles pareciam se mover para a frente e para trás em relação às estrelas do firmamento (movimento retrógrado). E isso ficou ainda mais evidente, quando no início do século XVII, os cientistas da época, como Galileu Galilei, começaram a utilizar o telescópio para a observação dos corpos celestes. Este dispositivo tecnológico marcou uma revolução no método da observação astronômica, permitindo que os astrônomos visualizassem maiores detalhes sobre o movimento dos planetas e suas características. Isso contribuiu para o colapso do geocentrismo e a necessidade de evoluir o modelo de sistema solar para um que pudesse explicar todas as observações existentes, que permitisse realizar previsões confiáveis sobre os movimentos planetários e que pudesse ser testado.

Esse momento da história é um bom exemplo para expor para o público geral que a ciência não é um empreendimento pronto ou mesmo exato, que fornece verdades absolutas. Podemos pensar na ciência como um processo, que evolui à medida que os seus métodos evoluem, ou seja, à medida que novas observações, novos experimentos, novas análises matemáticas e novas interpretações são realizadas. Qualquer estudo que é classificado como científico, envolve processos criteriosos com inúmeros testes e avaliações realizadas por profissionais da área. Para isso, investimentos em educação e no desenvolvimento de novas tecnologias são essenciais.

O professor pode estender o discurso sobre os princípios científicos e a evolução da ciência através da fala do narrador ou da criação de novos personagens conforme a realidade e o interesse de sua comunidade. Recomendamos que isso seja feito em mais de um esquete, para não sobrecarregar a audiência com muitos conceitos.

### *Esquete 2: O Modelo Heliocêntrico*

No esquete 2 o modelo heliocêntrico é apresentado como uma evolução do modelo geocêntrico. No início deste, o narrador pode expor de maneira introdutória tanto acontecimentos históricos quanto os benefícios da nova teoria para explicar o movimento dos corpos celestes.

Conforme já discutido, a representação do modelo de universo de Ptolomeu se manteve por quase 14 séculos. A partir de novas observações e do avanço desse método com o uso do telescópio, os astrônomos da antiguidade perceberam que era necessário evoluir essa teoria. No século XVI, Nicolau Copérnico propõe um modelo de sistema solar considerando que a Terra e os outros planetas orbitam o Sol. Além de orbitar o Sol, Copérnico considera que a Terra gira em torno de seu próprio eixo e que a Lua, apenas, orbita a Terra [28].

O modelo heliocêntrico permitiu explicar naturalmente o movimento retrógrado, a variação de brilho dos planetas e também mudanças observadas diariamente e sazonalmente nos céus. Além de corrigir as inconsistências do modelo geocêntrico, o sistema heliocêntrico de Copérnico possui uma característica muito valorizada por qualquer cientista, que é a sua simplicidade. Apesar da ciência parecer algo complexo e inacessível para a maioria das pessoas, os cientistas são atraídos e guiados pela simplicidade, simetria e beleza na modelagem de todos os aspectos da natureza.

Com a representação de Copérnico, o movimento do Sol e dos planetas em torno da Terra passa a ser apenas aparente, pois a Terra neste novo modelo orbita o Sol. Já na teoria de Ptolomeu esses movimentos são reais. Essa é uma diferença importante entre os dois modelos que pode favorecer a crença do público geral no geocentrismo, como no caso dos terraplanistas. Estas pessoas não aceitam ou não conseguem separar a realidade daquilo que enxergam, mesmo quando os argumentos apresentados a elas são robustos e bem fundamentados.

O movimento aparente está presente em nosso cotidiano. Se considerarmos que estamos sentados no banco de um ônibus em movimento, por exemplo, veremos o movimento aparente de pessoas, árvores e outros objetos que estão localizados do lado de fora do ônibus, pois como estamos parados no interior do veículo, temos a impressão de que tais objetos se distanciam ou se aproximam de nós, como se eles estivessem em movimento. Contudo, sabemos que uma árvore não pode se mover, uma vez que suas raízes a mantém fixa no chão. Esse conhecimento adicional nos permite concluir com certeza que é o ônibus que está em movimento, mesmo quando o seu deslocamento não é perceptível no interior do veículo. Se desenvolvermos um modelo que afirma que uma árvore está em movimento e o ônibus parado, este será rejeitado e até considerado absurdo por qualquer pessoa, pois ele não reproduz a realidade.

Mesmo apresentando exemplos simples como este, no intuito de fazer analogias, o público mais convicto em suas crenças pode argumentar que esse exemplo não é adequado, pois outra pessoa situada em um ponto de ônibus pode observar diretamente que é o ônibus que está em movimento e não uma árvore, enquanto que não é possível outra pessoa sair do planeta e ver que é a Terra que está em movimento e não o Sol. Mas a importância desse exemplo reside no fato de que movimentos aparentes existem e devem ser considerados diante da incompatibilidade do que é descrito por modelos e teorias, como o geocentrismo, e as observações realizadas sobre o comportamento da natureza, como o movimento relativo dos planetas. O cientista, neste caso, com seus instrumentos e métodos específicos e bem fundamentados, exerce o mesmo papel da pessoa situada no ponto de ônibus. Ele cria meios de observar melhor o comportamento da natureza e nos atualiza sobre isso, nos explicando o que pode estar acontecendo através de representações, descritas por modelos e teorias, cada vez mais aprimorados e consistentes com a realidade.

  A rejeição gratuita de ideias e avanços científicos sem argumentos e métodos bem fundamentados, em que são considerados dados observacionais, experimentais ou matemáticos, é conhecida como negacionismo científico. Este têm promovido o radicalismo, a desinformação e a desconfiança sobre temas importantes, como a eficiência das vacinas e as mudanças climáticas, podendo representar ameaças reais para a nossa sobrevivência em um futuro muito próximo, caso a ciência perca a credibilidade [30, 31]. Portanto, o problema hoje não é as pessoas questionarem sobre o que estão vendo e ouvindo, mas sim, não aceitarem a importância da ciência para o avanço da humanidade, mesmo diante de evidências concretas e convincentes.

  Trabalhar temas científicos em uma peça de teatro tem um grande potencial para atingir os familiares dos estudantes, pois além de tornar os conceitos acessíveis ao público geral através da arte, o conhecimento é transmitido pelo testemunho dos próprios estudantes. Isso pode tornar o conhecimento mais receptivo para as pessoas, pois o diálogo será estabelecido entre os membros das famílias e não a partir de pessoas desconhecidas. Para isso, é imprescindível que as pessoas, de maneira geral, entendam a importância e a necessidade dos investimentos em ciência, educação e tecnologia. É preciso acreditar na educação e passar a ver a escola como um espaço confiável, coletivo, acessível para todos, de transformação e busca pelo conhecimento.

  A importância de tais investimentos para a astronomia, por exemplo, irá favorecer a formação de novos profissionais para atuarem nesta área e possibilitar a construção de telescópios com maior alcance e poder de resolução, nos permitindo enxergar detalhes impossíveis de serem observados a olho nu. Isso pode nos auxiliar a prolongar a sobrevivência de nossa espécie no planeta a partir do monitoramento do Sol e sua influência sobre o clima na Terra e também de potenciais ameaças de meteoros que possam entrar em rota de colisão com o nosso planeta. Outra vantagem desses dispositivos é que eles permitem detectar uma variedade de luz que vai além daquela que enxergamos. A variedade de luz que conseguimos enxergar com nossos olhos e que nos permite detectar os objetos a nossa volta, suas cores e formas, é conhecida como espectro visível. Os telescópios modernos, situados na Terra e no espaço, nos permitem enxergar em toda a variedade de luz que existe na natureza, conhecida como espectro eletromagnético. O espectro visível é apenas uma parte do espectro eletromagnético. Este é apenas um exemplo da diversidade de temas e conceitos que podem ser discutidos em uma linguagem mais simples e acessível para o público geral no intuito de valorizar a ciência, o desenvolvimento de tecnologia e o ensino de ciências.

  Do ponto de vista educacional, estudantes do ensino fundamental e médio que tem oportunidade de participarem de atividades relacionadas com a astronomia, como a nossa peça de teatro, possuem maior probabilidade de seguir carreiras nas áreas de ciência e tecnologia, além de se manterem mais atualizados com relação às descobertas científicas [32].

A atuação dos estudantes-atores neste esquete foi semelhante ao anterior. Eles apenas entraram, se apresentaram e se posicionaram para representar a configuração do modelo heliocêntrico, com o Sol parado no centro do palco, os Planetas e a Terra girando em torno do mesmo através de órbitas circulares e a Lua orbitando a Terra.

*Esquete 3: As Três Leis de Kepler e a Lei da Gravitação Universal de Newton*

A necessidade de evoluir o modelo de sistema solar continua a ser apresentada no esquete 3. Apesar da revolução copernicana, com o modelo heliocêntrico, o matemático e astrônomo alemão Johannes Kepler percebeu inconsistências neste modelo a partir, principalmente, de uma extensa quantidade de dados observacionais do astrônomo dinamarquês Tycho Brahe [33]. O objetivo de Kepler com seus estudos foi encontrar uma descrição simples do movimento planetário que conciliasse a representação de Copérnico e os dados observacionais de Brahe.

Após anos de trabalho, Kepler percebeu que era necessário abandonar a ideia de Copérnico de órbitas planetárias circulares e que o Sol não ocupava o centro do sistema solar. Ele conseguiu descobrir as leis que regiam o movimento de todos os planetas conhecidos, incluindo a Terra, nomeadas como "as três leis de Kepler" [28].

O primeiro ato deste esquete foi para representar a descoberta da primeira lei de Kepler. Esta afirma que a órbita da Terra e de outros planetas em torno do Sol era ajustada pela curva geométrica de uma elipse. Uma vez que as órbitas dos planetas estão no mesmo plano que contém o Sol, a representação no palco do plano orbital da Terra foi feita simplesmente com a figura de uma elipse desenhada no chão. O Sol entra e ocupa um dos focos da elipse, marcado com um X. Logo em seguida, a Terra entra e gira em torno do Sol sobre o desenho da elipse. Os dois estabelecem um diálogo enquanto a Terra passa por diferentes posições, como o periélio e o afélio, para explicar o que são essas distâncias entre os dois corpos celestes. Em uma situação em que o palco seja suspenso, de maneira que a audiência não consiga enxergar o desenho da elipse, o narrador pode descrever a mesma como um círculo achatado ou alongado, fazendo gestos com as mãos, para facilitar o entendimento do público sobre a correção feita na órbita dos planetas por Kepler no modelo heliocêntrico.

A descoberta da segunda lei de Kepler foi representada pelas diferentes velocidades que a personagem Terra orbita o Sol. Quando a mesma passa pelo periélio, que define a região mais próxima do Sol, a Terra se move mais rapidamente, enquanto que na região mais afastada, definida pelo afélio, a Terra se move mais lentamente. Formalmente, essa lei é conhecida como a lei das áreas e afirma que a reta que liga o planeta ao Sol varre áreas iguais em tempos iguais.

Para a representação da terceira lei de Kepler, conhecida como lei harmônica, é interessante que outro estudante-ator entre no palco para fazer o papel de outro planeta orbitando o Sol. Esta lei estabelece que planetas com órbitas maiores se movem mais lentamente em torno do Sol e é muito interessante para mostrar para o público geral como tempo e distância são determinados astronomicamente no nosso sistema solar.

Tendo a Terra como referência, o ano terrestre, que é o período necessário para o nosso planeta completar sua órbita em torno do Sol, é a unidade de tempo. A unidade de distância é conhecida como Unidade Astronômica (UA) e corresponde à distância do Sol à Terra quando a mesma está localizada no ponto mais distante do Sol, ou seja, no afélio.

O professor deve avaliar a viabilidade de apresentar dados numéricos para deixar essa lei mais clara. Se considerarmos o planeta Mercúrio, que é um planeta que fica mais próximo do Sol do que a Terra, a sua distância em relação ao Sol, quando ele está na posição de afélio, é de 0,387 UA, e o tempo necessário para o mesmo completar uma órbita em torno do Sol é de 0,241 ano terrestre. Ou seja, Mercúrio leva em torno de 88 dias apenas para dar uma volta completa em torno do Sol enquanto que a Terra gasta 365 dias. Mercúrio é o planeta com movimento de translação mais rápido do sistema solar, pois este se encontra mais próximo do Sol e, portanto, possui a menor órbita. Para um planeta que está mais longe do Sol e possui uma órbita maior do que a da Terra, como Júpiter, por exemplo, sua distância em relação ao Sol será maior do que 1 UA e o seu tempo de órbita será maior do que 1 ano terrestre, conforme estabelecido pela lei harmônica [28].

As leis de Kepler foram obtidas por ele empiricamente. Este é outro conceito científico muito importante que pode ser trabalhado na peça.

Kepler conduziu um estudo empírico com base nos dados observacionais de Brahe. Isso significa que suas leis foram descobertas a partir da experiência de Brahe, ou seja, com base nas observações que Brahe fez do movimento dos corpos celestes. Se quaisquer dados empíricos sobre algum fenômeno da natureza forem obtidos com acurácia, através de processos e métodos que minimizam os erros das medidas ou observações, estes tornam-se inquestionáveis, pois os mesmos expressam de maneira precisa o comportamento da natureza. Por mais que um modelo teórico seja bonito e consistente, do ponto de vista matemático, por exemplo, este precisa reproduzir o que for obtido através de dados empíricos. Estes são resultados norteadores que mostram para os cientistas a validade ou a limitação de modelos teóricos. Os dados empíricos de Tycho Brahe permitiram que correções fossem introduzidas no modelo heliocêntrico e a descoberta das leis que regem o movimento dos planetas no sistema solar por Kepler.

Na última cena do esquete, a Terra apresenta algumas questões para provocar a audiência, como: o que impede os planetas de se distanciarem cada vez mais de você, Sol, ou de se encontrarem com você? Por que eu permaneço girando em torno de você por anos e anos de maneira aparentemente indefinida?

As respostas são dadas pelo Sol, que introduz a ideia de força gravitacional e como esta obedece a terceira lei de Newton, conhecida como lei da ação e reação. A representação de forças mútuas, com mesma intensidade, direção e sentidos contrários entre o Sol e a Terra, é esclarecida pelos estudantes-atores quando a Terra passa pela região do periélio. Eles mostram que a Terra só continua girando em torno do Sol porque eles dão as mãos, como mostrado na figura 1, e que esta força (gravitacional) continua existindo mesmo quando a Terra está a distâncias maiores do Sol. Eles deixam claro que quanto maior essa distância, menor a força e vice-versa. Como consequência disso, a Terra, assim como os outros planetas, diminui a sua velocidade de translação em conformidade com a segunda lei de Kepler.

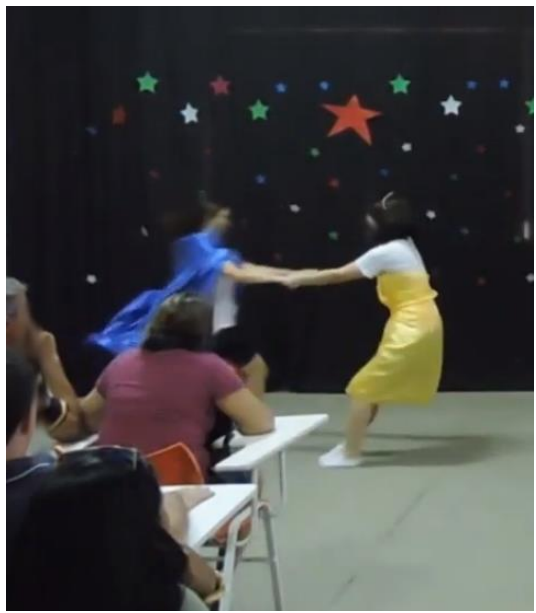

**Figura 1:** Foto mostrando as personagens Terra (azul) e Sol (amarelo) representando o conceito de força gravitacional e a lei de ação e reação quando a Terra passa pela região do periélio para justificar porque a Terra continua orbitando o Sol de maneira aparentemente indefinida.

O esquete é finalizado com o narrador dizendo que as leis de Kepler foram demonstradas matematicamente por Isaac Newton. Este é um exemplo da história da ciência que mostra a convergência de dois métodos científicos, a observação, que permitiu a obtenção de dados empíricos, e as análises matemáticas, que permitiram que um entendimento profundo sobre o movimento dos corpos celestes e suas causas fossem apresentados. O narrador contextualiza a relação direta da lei da gravitação universal com a queda de um objeto quando o soltamos, que é algo amplamente observado pelas pessoas no cotidiano.

### *Esquete 4: Os Dias e as Noites*

A existência dos dias e das noites é explicada pelos estudantes-atores através de um diálogo entre a Terra e o Sol. Eles explicam nossa unidade básica de tempo social, que é o dia solar de 24 horas, como uma consequência do movimento de rotação da Terra em torno de seu próprio eixo. Este movimento é representado principalmente pela rotação da cabeça do estudante em torno do eixo que passa pelo seu corpo. Mas o estudante realiza o movimento de maneira rígida, como se não fosse possível mexer o seu pescoço ou realizar qualquer outro movimento de torsão com o seu corpo.

Uma vez que a Terra translada em torno do Sol enquanto rotaciona, achamos interessante mencionar o que é o dia sideral e porquê este é aproximadamente 3,9 minutos menor do que o dia solar. Como não é possível posicionar a cortina de helanquinha, que representa as estrelas do firmamento, a uma distância muito maior daquela entre o Sol e a Terra no palco, a justificativa visual para a diferença de tempo entre os dias solar e sideral pode ser fornecida tendo como referência a própria audiência.

O dia solar ocorre quando um observador em nosso planeta, descrito pelos olhos da personagem Terra, estão na mesma posição em relação ao Sol. Isso representa a mesma hora do dia entre uma rotação e outra. Já o dia sideral fornece o período de uma rotação completa da Terra, que é o tempo que nosso planeta leva para retornar à mesma orientação no espaço em relação às estrelas distantes. Durante essa atuação é importante que o estudante-ator que for representar a Terra, além de manter o corpo rígido, também não realize movimentos laterais com os olhos. Para que a cena ganhe o significado desejado é necessário que o movimento dos olhos esteja condicionado apenas ao movimento da cabeça. Dessa forma, após uma rotação completa da cabeça do estudante-ator, a qual caracteriza um dia sideral, é fácil a audiência perceber que a direção dos olhos do mesmo, dada pela orientação do nariz, não estarão voltados diretamente para o Sol. Para que isso ocorra, ele precisará rotacionar um pouco mais, caracterizando o dia solar e explicando porque o mesmo é maior do que o dia sideral.

Um detalhe muito importante para a realização desta cena é considerar o movimento de rotação da Terra no mesmo sentido do seu movimento de translação em torno do Sol. Na nossa peça, como estamos representando um observador no Hemisfério Sul, a Terra e os outros planetas transladam em torno do Sol no sentido horário em todos os esquetes em que o movimento dos mesmos é considerado. Dessa forma, o movimento de rotação da Terra também deve ser realizado no sentido horário. Para ganhar um efeito visual ainda melhor, a personagem Terra pode realizar o seu movimento tendo como referência o alinhamento entre ela, o Sol e uma pessoa da audiência que fica atrás do Sol, por exemplo. Após uma rotação completa da Terra, esse alinhamento será visivelmente desfeito.

O esquete é finalizado com a Terra e o Sol explicando de maneira conclusiva, porque não é possível ver as estrelas do firmamento durante o dia claro.

Note como a linguagem corporal e artística dos estudantes-atores nesse esquete, que são características marcantes do teatro, são essenciais para materializar os conceitos científicos desejados e a comunicação dos mesmos para a audiência.

*Esquete 5: As Estações do Ano*

O narrador inicia o quinto esquete falando sobre a existência das quatro estações do ano e que as mesmas não ocorrem de maneira bem definida em quase todas as regiões do Brasil. Ele provoca a audiência com duas perguntas para nortear a atuação das personagens, uma sobre a causa das estações e a outra sobre a duração dos dias e das noites.

A ocorrência das estações é uma consequência da inclinação de 23,5º do eixo de rotação do nosso planeta em relação à sua órbita em torno do Sol, também conhecida como eclíptica. Durante o ano, a Terra ocupa quatro posições na eclíptica que definem o início de cada estação. Duas delas são conhecidas como equinócios e as outras duas como solstícios [33].

Os equinócios marcam o início do outono e da primavera. Nestas datas o dia claro e a noite possuem a mesma duração porque os raios solares incidem em nosso planeta perpendicularmente ao eixo de rotação da Terra, de maneira que a quantidade de luz solar que o planeta recebe é a mesma em qualquer região, como ilustrado na parte superior da figura 2. O equinócio de outono ocorre nos Hemisférios Norte e Sul aproximadamente nos dias 21 de setembro e 20 de março e o equinócio de primavera ou vernal ocorre nos dias 21 de março e 22 de setembro, respectivamente. Essas datas são aproximadas porque a duração real de um ano é de aproximadamente 365,25 dias e nós contabilizamos 365 dias. O tempo de 6 horas (0,25 dia) é compensado nos anos bissextos pelo acréscimo de 1 dia no calendário, conforme as regras estabelecidas no calendário gregoriano.

Os solstícios definem os dias claros e as noites mais longos do ano, correspondendo ao início do verão e do inverno, respectivamente. Durante estas datas a incidência dos raios solares não é a mesma nos Hemisférios Norte e Sul devido à inclinação do eixo de rotação da Terra. O solstício de verão é o dia que o Sol está mais alto no céu. Consequentemente, a região do polo correspondente fica mais exposta ao Sol. A incidência direta dos raios solares faz com que a temperatura se eleve originando climas mais quentes e úmidos. Enquanto isso, no polo oposto do planeta, há menor incidência de raios solares, pois o Sol está mais baixo no céu. Isso provoca quedas de temperatura e um clima mais frio e seco. A relação da posição do Sol no céu com as mudanças de temperatura é percebida por nós diariamente, pois as maiores temperaturas registradas durante o dia ocorrem por volta do meio-dia, que é o momento em que o Sol está mais alto no céu e os raios solares incidem diretamente sobre nós, formando regiões pequenas de sombra. Nos períodos da manhã e da tarde registramos temperaturas menores, pois o Sol está mais baixo no céu, de maneira que os raios solares incidem sobre nós de maneira inclinada, se espalhando por áreas maiores e formando grandes sombras. Isso é ilustrado na parte superior da figura 2 para uma região do Hemisfério Norte.

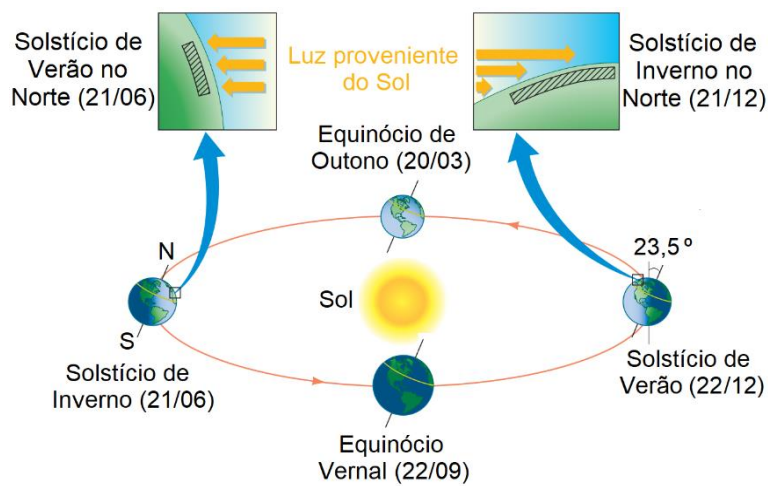

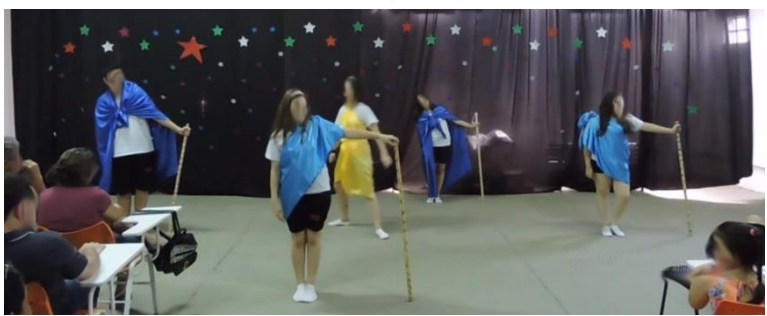

**Figura 2:** *Superior:* ilustração da Terra orbitando o Sol nas posições dos equinócios e dos solstícios para mostrar o início das quatro estações que ocorrem no Hemisfério Sul, como consequência da inclinação de 23,5º do eixo de rotação da Terra em relação ao plano de sua órbita. A circunferência feita na região Norte é para representar como os raios solares cobrem áreas diferentes nos pontos de solstício devido à essa inclinação, justificando as diferenças de duração entre os dias claros e as noites e das estações nos Hemisférios Norte (N) e Sul (S). Figura adaptada de [33]. *Inferior:* foto correspondente mostrando a atuação dos estudantes-atores.

Usualmente as pessoas acreditam que no inverno a Terra está mais distante do Sol do que no verão para justificar as diferenças de temperatura observadas nas duas estações. Se isso fosse verdade nós observaríamos uma mesma estação em todo o planeta e não diferentes estações no Hemisfério Norte e Sul, pois ambas as regiões estariam igualmente distantes ou próximas do Sol. Como discutido, o que ocorre realmente não são diferenças na distância entre o Sol e a Terra para definir as estações e controlar nosso clima, mas diferenças no tempo de exposição aos raios solares de uma região para outra devido à inclinação do eixo de rotação da Terra em relação à eclíptica. Portanto, quando alcançamos o solstício de verão no Hemisfério Sul, por volta de 22 de dezembro, o polo Sul estará mais exposto ao Sol fazendo com que os dias claros fiquem mais longos, as noites mais curtas e o clima mais quente. Neste mesmo período observa-se o solstício de inverno no Hemisfério Norte, com dias claros mais curtos, noites mais longas e climas mais frios.

Para representar os equinócios e os solstícios nas posições especificadas na ilustração da figura 2 foi necessário quatro estudantes-atores atuando como a Terra e um para atuar como o Sol. Eles foram nomeados na peça como Terra 1, 2, 3 e 4 para facilitar os ensaios. A Terra 1, ao fundo do palco, representou nosso planeta na posição do equinócio de outono, a Terra 2 à esquerda de quem olha para o palco ficou na posição de solstício de inverno, a Terra 3, à frente, na posição do equinócio vernal e a Terra 4, à direita, na posição do solstício de verão, como mostrado na foto apresentada na figura 2. A sequência das falas das personagens seguiu essa mesma ordem, no sentido anti-horário, conforme a ocorrência desses dias no calendário do ano.

Para representar o eixo de rotação da Terra e sua inclinação em relação a eclíptica, dada pelo plano do palco, os estudantes utilizaram cabos de vassoura, veja figura 2. Essa ideia surgiu nos ensaios pelos próprios estudantes para fornecer um efeito visual mais esclarecedor sobre isso. Contudo, durante a atuação, os estudantes também inclinaram o corpo na direção contrária e acabaram puxando consigo o bastão. Na foto apresentada na figura 2 somente a personagem Terra 2 se manteve na configuração desejada durante toda a cena. Observamos que, à medida que o tempo passa os estudantes-atores ficam cansados e acabam relaxando o corpo, pois eles se mantêm em uma única posição durante a cena. Isso faz com que o efeito visual da inclinação do eixo de rotação da Terra com o bastão acabe se perdendo. Para evitar este problema, aconselhamos que o bastão seja preso à frente do corpo dos estudantes-atores, através de um cinto ou na própria roupa, com a inclinação desejada. Dessa forma, basta eles ficarem de pé no palco que o efeito visual de inclinação do bastão será preservado durante toda a cena.

No final da cena o Sol e a Terra apresentam uma crítica justificada ao movimento terraplanista e o narrador valoriza a necessidade de investimentos em educação, ciência e tecnologia para conter as catástrofes ambientais, como o aquecimento global, causadas pela atividade humana.

### *Esquete 6: A Lua e o seu Lado Oculto*

A Lua é a principal personagem do sexto esquete da peça. O narrador inicia a cena provocando a audiência sobre a importância do nosso satélite natural e as missões Apolo que possibilitaram o homem a pisar na Lua. Essa provocação é importante porque muitas pessoas ainda acreditam que os pousos na Lua foram encenados [34].

Todo o conteúdo desta cena é desenvolvido através de um diálogo entre a Terra e a Lua. Além de fornecer inúmeras informações sobre a Terra, a Lua e o Sol, como suas massas e as distâncias relativas entre estes três corpos celestes, os estudantes-atores revisitaram o conceito de força gravitacional e discursaram sobre a influência da massa na mesma, em adição à sua dependência com a distância. Neste contexto, eles introduziram o buraco negro como um objeto supermassivo que até o Sol orbita e apresentaram o conceito de velocidade de escape para justificar porque a Lua orbita a Terra e não o Sol, uma vez que o Sol possui massa muito maior do que a Terra [35, 36].

O movimento dos estudantes-atores no palco é essencial para representar a sincronização que existe entre o movimento de rotação e o de translação da Lua, que faz com que a mesma apresente sempre a mesma face para o nosso planeta. Enquanto a Lua completa sua órbita em torno da Terra, esta última precisa rotacionar 27 vezes. Para que a cena não fique cansativa, aconselhamos que os estudantes realizem seus movimentos lentamente e que isso seja feito uma vez no início do diálogo, quando eles mencionarem a sincronização de seus movimentos, e seja repetido apenas quando a Lua finalizar os questionamentos sobre o porquê ela orbita a Terra. Essa repetição é necessária para que a Terra possa despertar a atenção do público para a sincronização dos movimentos e para introduzir a curiosidade sobre a existência ou não de um lado escuro na Lua. Para que a personagem Lua mantenha sempre a mesma face para a Terra é necessário que a mesma caminhe lateralmente. Marcações no palco podem ajudar os estudantes a realizar a sincronização de seus movimentos.

A Lua explica o que é dia lunar para convencer a Terra de que existe um lado oculto na mesma em relação aos observadores situados na Terra, mas que este não é necessariamente escuro. A Terra insiste na referência de um lado escuro ao mencionar o álbum de 1973 da banda de rock Pink Floyd intitulado "*The Dark Side of the Moon*" (O Lado Escuro da Lua), como um argumento que justificaria tal referência e, de certa forma, a crença popular sobre isso.

A música tem um papel muito importante no teatro para intensificar momentos dramáticos de uma cena ou aumentar o impacto geral da performance dos atores. Na nossa peça utilizamos a música *Brain Damage* [37] do álbum mencionado no final da cena para fornecer uma contextualização adicional para a mesma e criar uma conexão mais profunda com a audiência.

### *Esquete 7: As Fases da Lua e as Marés*

A Lua sempre fascinou a humanidade com suas fases, fornecendo mensalmente um espetáculo visual no céu noturno. Para explorar esse fenômeno na peça utilizamos quatro personagens Lua, de maneira semelhante ao que foi feito para as estações do ano no esquete 5. Cada personagem representou um quarto do ciclo sinódico da Lua, falando sobre a sua posição em relação ao Sol e a Terra e as fases nova, crescente, cheia e minguante.

Através de um diálogo descontraído e divertido entre o Sol, a Lua e a Terra, os estudantes-atores explicaram a diferença entre mês sinódico e sideral, em analogia aos dias solar e sideral. O roteiro da peça foi elaborado para revisitar ou resgatar conceitos já discutidos em esquetes anteriores sempre que possível. Nosso objetivo com isso foi familiarizar a audiência com os conceitos e a linguagem científica utilizada ao longo do espetáculo para a explicação dos fenômenos.

Além do teatro permitir trabalhar o cognitivo e a razão com a linguagem corporal dos estudantes-atores, é possível também explorar a cultura popular, como nós fizemos através de uma das lendas de folclore mais conhecidas do mundo, o lobisomem. A transformação de um homem em lobo em noites de Lua Cheia pode fornecer um contexto muito interessante para despertar um interesse ainda maior da audiência pela cena, devido à familiaridade que as pessoas têm com a história do lobisomem. O discurso científico sobre qualquer assunto não precisa ser rígido e completamente técnico, o professor pode explorar o ambiente e as possibilidades oferecidas pelo teatro para transformá-lo em algo mais acessível e agradável.

A introdução das marés foi feita naturalmente logo após a explicação das fases da Lua, com o Sol perguntando para a Terra se a mesma também possuía fases. As perguntas norteadoras feitas pelas personagens ao longo de toda a peça foram preparadas com o objetivo de estabelecer relações daquilo que havia acabado de ser discutido ou explicado com os novos conceitos que estavam sendo introduzidos. Com essa abordagem tentamos criar contextos para facilitar a assimilação dos novos conceitos e o estabelecimento de correlações pela própria audiência.

Aproveitamos o tema das marés para introduzir tópicos de educação ambiental. Ao falar sobre a proporção de água salgada e doce que existe em nosso planeta, chamamos a atenção da audiência para a importância de evitar a poluição de lagos e rios.

Para a explicação da causa das marés revisitamos o conceito de força gravitacional apresentando números que evidenciam que a força que o Sol exerce na superfície da Terra é muito maior que a força exercida pela Lua. Essa discussão é importante para deixar claro para o público que são as variações do campo gravitacional que provocam as marés e não necessariamente a intensidade dessas forças. Apesar das marés serem um resultado direto da influência do campo gravitacional do Sol e da Lua, a contribuição da Lua é maior para a ocorrência do fenômeno porque a mesma está bem mais próxima da Terra do que o Sol. Devido à essa proximidade a variação do campo gravitacional lunar em toda a Terra é consideravelmente maior do que a do campo solar [38], justificando a relação das marés com as fases da Lua.

A cena é encerrada com a valorização da ciência quando a Terra fala sobre a existência de marés terrestres e atmosféricas e que as mesmas só podem ser detectadas por instrumentos científicos muito sensíveis.

### *Esquete 8: Eclipses Lunares e Solares*

No último esquete da peça apresentamos os eclipses, que estão entre os fenômenos mais espetaculares da natureza e conhecidos por todos.

O narrador introduz o tema com três perguntas norteadoras sobre a causa dos eclipses, sua frequência e o tipo de eclipses que existem. Estas são respondidas ao longo da cena. Assim como no esquete anterior, toda a trama foi desenvolvida através de um diálogo entre o Sol, a Lua e a Terra.

Para que o eclipse lunar não fosse confundido com a sombra da própria Lua formada em sua face oposta ao Sol, as personagens Terra e Lua apresentaram um contexto para diferenciar esta da sombra projetada na Lua pela Terra e introduzem o tema.

Em seguida o Sol participa da cena passando a impressão de que ele é o protagonista da narrativa sobre os eclipses, pois os raios solares são bloqueados pela Terra, obscurecendo a Lua em um eclipse lunar, e bloqueados pela Lua, tendo como resultado a projeção de uma sombra na superfície da Terra durante um eclipse solar. Contudo, a Lua explica que o seu movimento orbital em torno da Terra exerce um papel essencial na ocorrência dos tipos de eclipses que observamos e na frequência dos mesmos, devido à inclinação de sua órbita de aproximadamente 5º, em relação ao plano da eclíptica, e a mudança de direção dessa inclinação em relação ao Sol. Dessa forma, a Lua toma a narrativa para si para a explicação detalhada dos fenômenos com o auxílio de um colar de isopor utilizado pela Terra para representar sua órbita em torno da mesma.

O colar de isopor é um artefato essencial para mostrar para a audiência que a inclinação da órbita lunar em relação à eclíptica faz com que não haja um alinhamento perfeito entre o Sol, a Lua e a Terra todas as vezes que a Lua está em suas fases Nova e Cheia. Isto é ilustrado na parte superior da figura 3. Além da inclinação da órbita lunar, a personagem Terra consegue auxiliar a Lua em sua explicação para mostrar que a frequência de ocorrência dos eclipses é afetada pela mudança da direção dessa inclinação em relação ao Sol, enquanto ambas orbitam o Sol. Esse efeito é demonstrado visualmente com a Terra realizando um movimento de bamboleio com o colar de isopor em torno de seu pescoço. Duas fotos dessa cena são mostradas na parte inferior da figura 3. Na foto à esquerda é capturado o momento em que a Lua está na fase Nova, mas os três corpos celestes não estão perfeitamente alinhados, de maneira que o eclipse solar não ocorre. Nesta configuração a Terra vê a Lua abaixo do Sol. Note que o colar de isopor (órbita lunar) está visivelmente inclinado em relação ao chão do palco (eclíptica) nesta imagem. Na foto à direita é mostrado o momento em que a Lua cruza o plano da eclíptica quando a mesma está na fase Cheia. Como os três corpos celestes estão perfeitamente alinhados os estudantes-atores explicam a ocorrência de um eclipse lunar total, com a personagem Lua se escondendo atrás da Terra para não ser vista pelo Sol. Os corpos celestes nesta cena são descritos pelas cabeças dos estudantes-atores.

Os estudantes-atores explicaram a cor avermelhada aparente que a Lua adquiri durante um eclipse lunar total e aproveitaram o contexto para falar sobre as cores do céu durante o dia e durante o pôr do Sol. Adicionalmente, eles mencionaram mitos referentes a guerras, maus presságios ou catástrofes naturais iminentes, criados pela humanidade na antiguidade para justificar o fenômeno.

Esse momento foi oportuno para valorizar o desenvolvimento científico, mostrando que a ciência nos fornece um conhecimento organizado e racional do mundo que vivemos, frente à nossa relação emocional decorrente da observação de fenômenos naturais, como os eclipses, que podem culminar no obscurecimento de nossos julgamentos e induzir comportamentos e atitudes indesejados diante da sociedade.

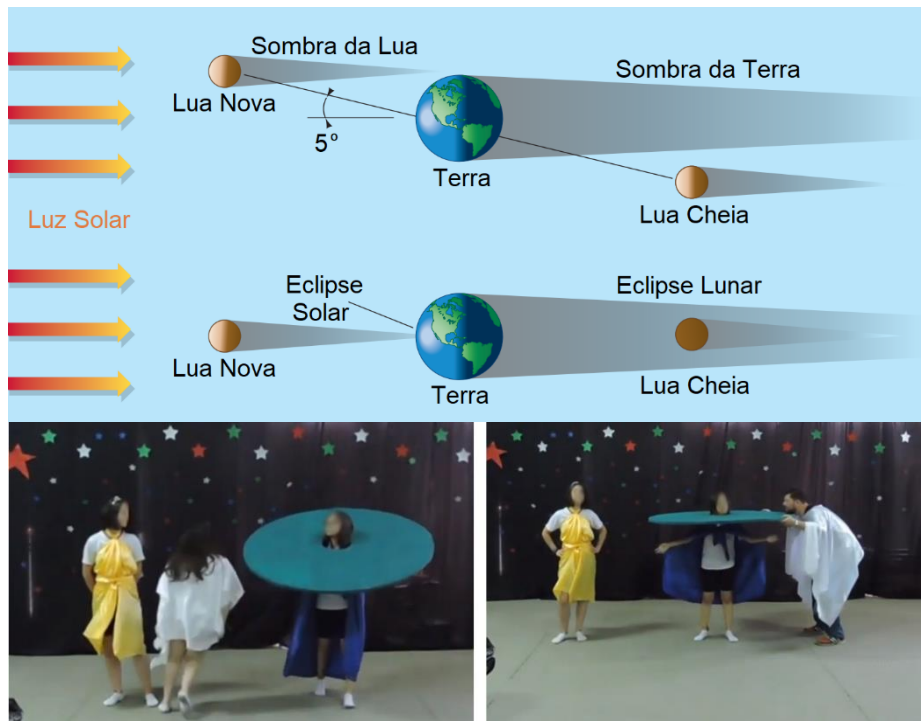

**Figura 3:** *Superior:* ilustração mostrando que os eclipses não ocorrem em todas as Luas Nova e Cheia devido à inclinação de aproximadamente 5º da órbita da Lua em relação à eclíptica. Para as configurações em que o alinhamento entre o Sol, a Lua e a Terra é perfeito, os eclipses solares na Lua Nova e os lunares na Lua Cheia podem ser observados. Figura adaptada de [33]. *Inferior:* fotos correspondes mostrando a atuação dos estudantes-atores, com a Terra utilizando um colar de isopor para representar a órbita lunar, o Sol e a Lua nas cores amarelo e branco, respectivamente.

Os estudantes-atores explicaram como os eclipses totais e parciais ocorrem e as regiões de sombra formadas na superfície da Terra durante um eclipse solar, a umbra e a penumbra. Em seguida a Lua fala sobre a frequência anual de eclipses solares e lunares de maneira estimada. Diante da falta de rigor nas estimativas feitas pela Lua, a Terra pergunta se existe um período mais preciso para prever a ocorrência dos mesmos, como o mês sinódico, em que as fases da Lua se repetem periodicamente e de maneira bem definida.

Utilizamos este contexto para introduzir o ciclo de Saros, referente à periodicidade de ocorrência dos eclipses. Este período é conhecido desde a antiguidade, pelo menos há 500 anos antes da era cristã, e é pouco familiar entre os estudantes e o público geral. Os povos babilônios conseguiram determinar o ciclo de Saros revelando a acurácia de suas observações astronômicas, cujos diários foram essenciais para os gregos construírem a máquina de Anticífera (*Antikythera*) no século I antes de Cristo. Este dispositivo mecânico é provavelmente o computador analógico e planetário mais antigo que se conhece e era utilizado para prever a posição de corpos celestes e os eclipses. Este ciclo também pode ter inspirado a construção do monumento megalítico pré-histórico localizado na Inglaterra, conhecido como Stonehenge. Apesar de sua função principal ainda estar em discussão entre os especialistas, a teoria mais aceita é que o mesmo tinha como principal propósito prever eclipses [39, 40].

A simples apresentação do ciclo de Saros pode ser útil para mostrar que o conhecimento científico evoluiu ao longo dos séculos juntamente com os seus conceitos, técnicas e o desenvolvimento de novas tecnologias.

O esquete é finalizado com uma discussão sobre a dificuldade de a ciência prosperar em uma sociedade onde o negacionismo científico e a manipulação de informação prevalecem. Contudo, uma possível solução é apresentada para conscientizar a audiência sobre a necessidade de investimentos em uma educação igualitária, laica e pública de qualidade e a valorização das escolas e universidades públicas.

Ao final, saem do palco a Lua, a Terra e o Sol e em seguida entram todos e agradecem ao público, como mostrado na figura 4.

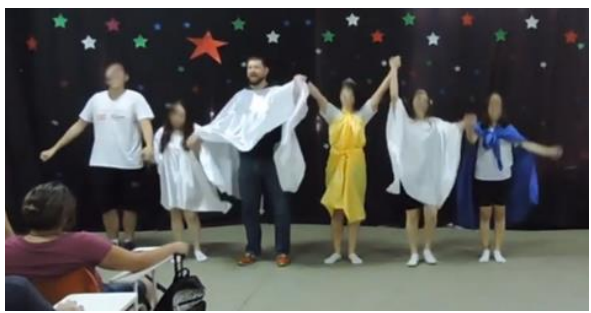

**Figura 4:** Foto mostrando o final do espetáculo com o professor responsável pela peça no centro e os cinco estudantes-atores agradecendo a audiência.

## 4. Considerações Finais

A avaliação do uso do teatro como metodologia para comunicar a ciência para o público geral e para auxiliar os estudantes envolvidos na peça a entender melhor os conceitos propostos foi feita pelo professor de maneira qualitativa, focando principalmente na abordagem e no comportamento dos estudantes [41, 42].

A valorização da improvisação nas falas dos estudantes durante os ensaios forneceu maior liberdade aos mesmos para expressar os conceitos científicos com suas próprias palavras. Sempre que um novo conjunto de definições era apresentado pelo professor, esperava-se que os estudantes não utilizassem a terminologia canônica da Física corretamente o tempo todo. Neste contexto, o professor trabalhou ativamente para buscar significados e conteúdos ocultos que ajudassem a melhorar a atuação dos estudantes na peça.

Essa abordagem nos propiciou uma descoberta muito importante para a divulgação científica e o ensino de Física ou de ciências, de maneira geral. Se o professor for capaz de correlacionar os conceitos e a terminologia científicos com o que os estudantes estão expressando, palavras e termos técnicos podem ser substituídos em favor de uma maior clareza de linguagem, conduzindo a uma maior clareza dos próprios conceitos. E isso pôde ser percebido pela satisfação da plateia, que parecia se encantar e se divertir com o conteúdo científico da peça apresentado pelos estudantes-atores em uma linguagem acessível para o público geral, protagonizada por eles mesmos.

Apesar dos benefícios mencionados é necessário que o professor fique atento para que a improvisação de um estudante-ator não bloqueie a atuação dos outros, seja por falta de entendimento do que está sendo exposto ou pelo desejo daquele estudante de controlar a forma como as falas dos outros devem prosseguir. Isso pode conduzir a desentendimentos e frustrações e, consequentemente, comprometer o bom andamento da peça. Trabalhar a coletividade e a colaboratividade entre os estudantes é fundamental para criar uma atmosfera agradável e produtiva.

Não podemos deixar de mencionar que a expressão corporal e facial, que são características marcantes do teatro, foram grandes facilitadores para os estudantes-atores materializarem os conceitos científicos, fazendo com que os fenômenos discutidos e as características de cada personagem parecessem reais.

Além dos benefícios de comunicar a ciência de uma maneira descontraída e divertida para a comunidade escolar, podemos concluir que a experiência fornecida aos estudantes através do teatro permitiu romper a relação professor-estudante usualmente vivida em sala de aula, para criar uma interação mais significativa, memorável e colaborativa no processo ensino-aprendizagem. Finalizamos com uma mensagem de um dos estudantes-atores sobre o seu envolvimento com o trabalho desenvolvido para salientar a importância de se implementar o pluralismo metodológico no ensino de ciências:

*"Apesar de não gostar muito da matéria [Física] em si, não tive experiências ruins, apenas adquiri mais vontade de estudar. Os encontros do teatro me ensinaram algumas virtudes, amor e respeito ao semelhante, paciência, concentração e empenho."*

## 5. Material Suplementar

No material suplementar apresentamos o roteiro autoral completo da peça "Sistema Maluco" com sugestões de expressões e movimentos para tornar os conceitos físicos visualmente mais acessíveis para o público. É importante salientar que, apesar do roteiro apresentar uma descrição detalhada de conceitos e fenômenos, nosso objetivo foi propor um material de ensino norteador para mostrar ao professor novas possibilidades para o ensino de ciências, de maneira que o mesmo pode ser adaptado e modificado conforme a realidade da comunidade escolar.

## Agradecimentos

## Referências

# "Sistema Maluco"
por Ederaldo Bueno de Macedo Junior e James Alves de Souza

# Introdução

**Narrador:** Olá senhoras e senhores, sejam todos muito bem-vindos ao nosso teatro. Nós apresentaremos a peça intitulada "Sistema Maluco" para fazer referência ao nosso maravilhoso e misterioso Sistema Solar. Com as nossas encenações acerca da Física vamos apresentar, de maneira lúdica e divertida, como alguns dos desafios enfrentados pela ciência foram resolvidos no campo da Astronomia.

A ciência é uma atividade que vem sendo desenvolvida e aprimorada pelos cientistas desde os primórdios da era cristã. A astronomia é uma área da ciência que estuda os corpos celestes, como o Sol, a Lua, os planetas e todos os fenômenos relacionados aos mesmos. A observação do firmamento foi talvez a atividade mais prazerosa das civilizações de eras passadas e é ainda praticada por todos nós atualmente. Afinal, quem nunca se encantou com o céu noturno!

Contudo, nossas observações a olho nu e nossas curiosidades são limitadas e podem nos levar a crenças que não estão necessariamente relacionadas com a realidade do nosso universo. A observação feita pelos cientistas é uma atividade extremamente profissional, sendo mais cuidadosa e criteriosa para evitar que fake news, interesses pessoais de políticos e pessoas que querem lucrar indiscriminadamente sejam utilizados para enganar as pessoas e contribuir para a destruição da natureza. Através de métodos muito bem fundamentados e instrumentos específicos, eles nos revelam inúmeros mistérios sobre o universo, que nos permitem entender e apreciar vários fenômenos observados no nosso dia a dia, como o porquê de existir dias e noites, as quatro estações do ano, as fases da lua, as marés, os eclipses, entre outros.

Convidamos a todos e todas para permanecerem conosco nessa incrível jornada para entender esses fenômenos e muito mais com a nossa peça.

**Esquete 1: O Modelo Geocêntrico**

**Narrador:** Para entendermos melhor os fenômenos que serão discutidos, consideraremos um observador no Hemisfério Sul da Terra, que é onde o Brasil está localizado. As estrelas do firmamento que caracterizam o céu profundo serão representadas pela cortina presente no palco.

O primeiro sistema planetário concebido para explicar o movimento dos corpos celestes, considerados como esféricos, foi proposto pelo astrônomo e geógrafo grego Cláudio Ptolomeu no século II da era cristã. Este é conhecido como modelo geocêntrico ou geocentrismo. Apesar do movimento atual de pessoas que acreditam equivocadamente que a Terra é plana, conhecido como terraplanismo, o reconhecimento do formato esférico da Terra data de 300 anos antes de Cristo.

Mas como funciona o modelo geocêntrico de Ptolomeu?

- Entra a personagem Terra, se posiciona no centro do palco e se apresenta:

**Terra:** Eu sou a Terra e ocupo o centro do Universo no modelo geocêntrico.

- Em seguida entram as personagens Lua, Sol e Planetas girando no sentido horário ao redor da Terra em órbitas circulares e se apresentam:

**Lua:** Eu sou a Lua e giro em torno da Terra.

**Sol:** Eu sou o Sol e giro em torno da Terra.

**Planetas:** Eu represento os Planetas que também giram em torno da Terra nesse modelo.

- A Terra olha para a cortina do palco e gesticula dizendo:

**Terra:** E no firmamento encontram-se as estrelas fixas.

- Saem do palco os Planetas, o Sol e a Lua nesta ordem, girando no sentido horário em torno da Terra e por último sai a Terra.

**Esquete 2: O Modelo Heliocêntrico**

**Narrador:** Como vocês puderam perceber, no modelo geocêntrico a Terra ocupa o centro do universo e a Lua, o Sol e os Planetas orbitam a mesma. Com esse modelo é possível explicar e até prever o movimento aparente do Sol e da Lua no céu. Contudo, os cientistas constataram, a partir de observações cuidadosas ao longo dos anos, que não era possível explicar, com o geocentrismo, o movimento dos 5 planetas conhecidos naquela época: Mercúrio, Vênus, Marte, Júpiter e Saturno. Portanto, foi preciso propor um novo modelo de sistema planetário.

Isso aconteceu após mil e quatrocentos anos, já no século XVI, que foi quando o astrônomo e matemático polonês Nicolau Copérnico propôs o sistema heliocêntrico. Essa ideia já havia sido proposta na Grécia antiga por Aristarco de Samos no século III antes de Cristo, mas foi rejeitada por outros cientistas gregos.

- Entra a personagem Sol, se posiciona no centro do palco e diz:

**Sol:** Eu sou o Sol e ocupo o centro do Universo no modelo heliocêntrico.

- Entra a personagem Terra, gira no sentido horário ao redor do Sol, e diz:

**Terra:** Eu sou a Terra, além de girar em torno do Sol em uma órbita circular, eu giro em torno do meu próprio eixo em um movimento de rotação nesse novo modelo.

- Entra a personagem Lua, girando no sentido horário ao redor da Terra e diz:

**Lua:** Eu sou a Lua e eu sou o único corpo celeste que gira em torno da Terra no modelo heliocêntrico.

- Entra a personagem que representa os planetas girando no sentido horário em torno do Sol e diz:
**Planetas:** Eu sou os Planetas que também giram em torno do Sol em órbitas circulares.
**Sol:** E no firmamento encontram-se as estrelas fixas.
- Saem do palco os Planetas, a Lua e a Terra nesta ordem, girando no sentido horário em torno do Sol e por último sai o Sol.

**Esquete 3: As Três Leis de Kepler e a Lei da Gravitação Universal de Newton**

**Narrador:** A evolução do método da observação na astronomia permitiu que os cientistas concluíssem que o modelo heliocêntrico é mais consistente com a realidade do que o modelo geocêntrico. Com o novo modelo foi possível entender, de maneira inquestionável, que a Terra não está no centro do universo, que todos os planetas do sistema solar orbitam o Sol e que qualquer movimento aparente das estrelas do firmamento é resultado do movimento de rotação da Terra, uma vez que estas estrelas estão muito distantes do nosso planeta. Contudo, novas observações permitiram que os cientistas aprimorassem também o modelo heliocêntrico.

No início do século XVII, o astrônomo, astrólogo e matemático alemão Johannes Kepler, analisando os dados observacionais do astrônomo dinamarquês Tycho Brahe, percebeu que as órbitas dos planetas em torno do Sol não eram circulares e o Sol não ocupava o centro do Universo. Após anos de estudo, ele conseguiu descobrir leis precisas que regem o movimento de todos os planetas conhecidos, assim como o de cometas, nomeadas como "as três leis de Kepler".
- O Sol entra e chama a atenção do público para a curva geométrica que está desenhada no chão do palco:
**Sol:** Eu sou o Sol e Kepler percebeu que eu não ocupo o centro do universo. A posição que estou agora é um dos focos de uma figura geométrica conhecida como elipse. Ela está desenhada no chão a minha volta e parece uma circunferência achatada. Mas vocês sabem o que esta elipse representa?
- A Terra entra e gira no sentido horário ao redor do Sol sobre a curva da elipse e diz:
**Terra:** Eu sou a Terra e esta elipse é a trajetória que eu percorro em torno do Sol. Ela também é conhecida como eclíptica.
- Os planetas entram e dizem:
**Planetas:** Olha Sol, assim como a Terra, nós também giramos em torno de você através de órbitas elípticas. E isso ocorre se estivermos mais próximos de você, em órbitas menores, ou mais distante de você do que a Terra, em órbitas maiores. Isso não é legal?
**Sol:** Sim! A primeira lei de Kepler afirma exatamente isso: a órbita da Terra e de outros planetas em torno de mim, o Sol, é ajustada pela curva geométrica de uma elipse.
- Os Planetas saem e a Terra se move até a posição do periélio, para e diz:
**Terra:** Sol, agora que você não está mais no centro do universo eu consigo ficar mais próxima e mais distante de você. Você sabia disso?

**Sol:** Sim Terra, a posição que você está parada se chama Periélio e esta é a distância mais próxima que ficamos um do outro durante o ano.

- A Terra se move até a posição do afélio, para e diz:

**Terra:** Isso é muito interessante Sol, e quanto à posição mais distante de você, ela também tem um nome?

**Sol:** Sim, esta se chama Afélio e é o período do ano que eu sinto mais saudades de você.

**Terra:** Sol, uma coisa muito interessante que eu estou observando é o seguinte: quando eu passo pelo periélio a minha velocidade é maior do que quando eu passo pelo afélio. Por que isso acontece?

**Sol:** Isso ocorre porque a área que você percorre em torno de mim na eclíptica deve ser a mesma em tempos iguais. Essa é uma lei da natureza e também foi descoberta por Kepler, sendo chamada de segunda lei de Kepler.

- A Terra representa a segunda lei de Kepler enquanto orbita o Sol, aumentando e diminuindo a sua velocidade e apontando a área varrida com as mãos.

**Terra:** Deixa eu ver se entendi! O meu movimento é variado porque ao me aproximar de você eu varrerei áreas menores, pois nossa distância é menor, e ao me distanciar de você varrerei áreas maiores, porque nossa distância é maior.

**Sol:** Muito bem Terra, mas não se esqueça que você deve considerar os tempos iguais.

**Terra:** Então, para varrer a mesma área em tempos iguais eu preciso mudar a minha velocidade, sendo mais rápida na região que eu percorro áreas menores e mais lenta na região que percorro áreas maiores. Que incrível!

**Sol:** Isso mesmo. Opa, é melhorar você se apressar, pois está chegando no periélio.

**Terra:** Sol, eu tenho outra pergunta?

**Sol:** Pois não Terra, eu estou aqui para iluminar o seu caminho!

**Terra:** Como podemos medir o tempo e a distância em nosso sistema planetário?

**Sol:** Isso é muito simples Terra, nós utilizamos você como referência. O tempo necessário para você completar sua órbita em torno de mim é a unidade de tempo, chamada de ano terrestre e corresponde a 365,25 dias.

**Terra:** Sim, to ligada!

**Sol:** Os bichos que moram em você aproximam este tempo para 365 dias e compensam a diferença nos anos bissextos, que possuem 366 dias.

- A Terra fala com estranheza:

**Terra:** Bichos? Ah sim, os seres humanos.

**Sol:** Isso, eu esqueci de como eles são chamados (risos).

**Terra:** Entendi, e quanto à distância?

**Sol:** Já as distâncias entre os corpos celestes são medidas pela Unidade Astronômica (UA). Esta equivale à distância entre você e eu quando você está localizada no afélio, aproximadamente 150 milhões de quilômetros.

**Terra:** Nossa! Essa é uma distância muito grande.

**Sol:** Sim Terra, não é ali!

**Terra:** Sol, então, de acordo com esse sistema de medida, os planetas que estão mais próximos de você do que eu, como Mercúrio e Vênus, completam uma volta em menos de um ano e estão à distância de menos de 1 UA de você, certo?
**Sol:** Isso mesmo Terra, pois você é a nossa referência.
- Entra o planeta Mercúrio percorrendo uma órbita elíptica entre o Sol e a Terra em alta velocidade dizendo:
**Mercúrio:** Zum zum zum! Eu sou Mercúrio o planeta mais rápido do sistema solar. Estou a 0,387 UA de distância do Sol e consigo completar uma volta em torno dele em apenas 0,241 ano, ou 88 dias. Ninguém me pega!
- Sai o planeta Mercúrio.
**Terra:** Minha nossa, além de Mercúrio ser o planeta mais rápido do sistema solar, ele também deve ser o mais maluco!
**Sol:** Risos... esse é o nosso ligeirinho. Da mesma forma Terra, se você olhar para os planetas mais distantes de mim do que você, como Marte, Júpiter e outros, eles estarão a mais de 1 UA de mim e demorarão mais de 1 ano para completar suas órbitas em torno de mim.
- Uma voz nos bastidores surge para representar Júpiter falando a uma distância muito grande.
**Júpiter:** Oi Terra, eu sou Júpiter e estou a uma distância do Sol de 5,203 UA e demoro quase 12 anos para completar minha órbita.
- A Terra coloca a mão sobre os olhos e observa na direção que veio a voz de Júpiter e diz:
**Terra:** Oi Júpiter, minha nossa, você está muito longe mesmo, olhando daqui você é apenas um pontinho brilhante no céu noturno. Tchau!
**Terra:** Sol, a partir dessas observações eu consegui chegar à conclusão de que planetas com órbitas maiores se movem mais lentamente em torno de você. É isso mesmo?
**Sol:** Perfeitamente Terra! Essa observação é precisamente o que estabelece a terceira lei de Kepler.
**Terra:** Uau que interessante! Sol, mas eu ainda tenho uma dúvida com relação a órbita dos planetas.
**Sol:** Pode falar Terra, mas espero que seja a última, porque esta cena já está muito longa.
**Terra:** O que impede os planetas de se distanciarem cada vez mais de você ou de se encontrarem com você? Ou seja, por que eu permaneço girando em torno de você, por anos e anos, de maneira aparentemente indefinida?
**Sol:** Isso é por causa da força de atração que existe entre nós, chamada de força gravitacional. A força com que eu atraio você possui a mesma intensidade, direção e sentido contrário da força com que você me atrai. Contudo, essas forças dependem da distância, ou seja, quanto mais longe de mim você estiver, menor será nossa força de atração.
**Terra:** Que legal Sol! Então é por isso que eu aumento a minha velocidade quando estou mais próxima de você na região do periélio, porque nossa força de atração é maior.

- A Terra dá as mãos para o Sol quando passa pelo periélio.
**Sol:** Sensacional Terra! Acho que depois de nossas explicações, muitas crianças e jovens que estão nos assistindo vão querer seguir uma carreira científica.
**Terra:** Espero que sim Sol, pois é preciso mais cientistas para ajudarem a cuidar melhor de mim.
- Terra e Sol saem do palco.
**Narrador:** As leis de Kepler foram demonstradas matematicamente pelo matemático, físico e astrônomo britânico Isaac Newton no final do século XVII, em torno de 1687. As análises matemáticas também é um método científico. Através deste, Newton descobriu a lei da gravitação universal, que explica não só a força de atração entre os corpos celestes, mas também porque um objeto cai no chão quando o soltamos.

**Esquete 4: Os Dias e as Noites**

**Narrador:** Através da ciência e da evolução da tecnologia foi possível melhorar consideravelmente a nossa representação do sistema solar. Isso nos forneceu um entendimento profundo sobre a natureza para explicar e até prever não só o movimento dos corpos celestes, mas também inúmeros outros fenômenos, como a existência dos dias e das noites e porquê não enxergamos as estrelas durante o dia claro.
- O Sol entra e se posiciona em um dos focos da eclíptica.
- A Terra entra realizando o movimento de translação no sentido horário em torno do Sol e gira em torno de si mesma, também no sentido horário, para representar o movimento de rotação, e pergunta:
**Terra:** Sol você sabe por que existem os dias e as noites em mim?
**Sol:** Oi Terra, eu nem sei o que é isso!
**Terra:** É simples Sol, os dias ocorrem quando eu vejo você e tudo fica iluminado. Já as noites ocorrem quando eu estou de costas para você e não consigo ver o seu brilho, tudo fica escuro.
**Sol:** Minha nossa Terra, eu pensava que eu iluminava tudo que estava a minha volta. Mas como isso acontece?
**Terra:** Enquanto eu giro em torno de você, eu também giro em torno do meu próprio eixo. Este é chamado de movimento de rotação. Por causa disso, eu fico alternando periodicamente em intervalos de tempo claro e escuro, ou seja, dia e noite.
**Sol:** Periodicamente? Ah sim, em intervalos repetidos, dia e noite, dia e noite, ... (risos)! E qual é o período deste movimento?
- A Terra se posiciona em frente ao Sol, olhando para ele diretamente nos olhos e realiza os movimentos de translação e rotação mantendo o corpo e a cabeça rígidos. A direção do olhar da Terra deve ficar alinhada com o seu nariz. Para conseguir o efeito desejado, a personagem não pode realizar movimentos laterais com os olhos ou com o pescoço.

**Terra:** Para retornar a esta posição em que os meus olhos estão alinhados com os seus eu demoro 24 horas. Isso determina o mesmo horário do dia anterior, sendo por isso, chamado de dia solar.
**Sol:** Espera aí Terra! Por que dia solar, existe outro tipo de dia?
**Terra:** Sim, também existe o dia sideral. Este determina o tempo para eu realizar uma rotação completa em torno do meu eixo.
**Sol:** Mas por que essa diferença? Isso tudo não é a mesma coisa?
**Terra:** Não! O dia solar é medido em relação a você, quando os meus olhos encontram os seus, enquanto que o dia sideral é medido em relação à minha orientação no espaço.
**Sol:** Orientação no espaço? O que isso significa?
**Terra:** É a minha orientação em relação as estrelas do firmamento. Como eu também estou realizando um movimento de translação em torno de você, eu preciso girar um pouquinho mais para conseguir ver os seus olhos e estabelecer o mesmo horário do dia anterior. Vou girar e transladar bem devagar para você perceber a diferença entre os dois dias.
- A Terra pode estender a mão para a frente para auxiliar a audiência a perceber quando ela completa o seu movimento de rotação antes de ficar completamente alinhada com o Sol.
**Sol:** Que interessante Terra! A combinação dos seus dois movimentos, rotação e translação, é o que causa esta diferença de tempo. Você saberia dizer quanto é essa diferença em minutos?
**Terra:** É claro! O dia solar é maior do que o dia sideral em aproximadamente 3,9 minutos.
**Sol:** Nossa, é o tempo de um cafezinho!
**Terra:** Sol, agora que esclarecemos como ocorrem os dias e as noites, acho que fica fácil explicar porque eu não consigo ver as estrelas do firmamento durante o dia claro.
**Sol:** Com certeza Terra! Sendo eu uma estrela tão bela, tão intensa e estando mais próxima de você do que as outras, o meu brilho ofusca as demais estrelas do universo durante o dia claro.
**Terra:** Perfeitamente Sol e como eu não vejo o seu brilho durante a noite, eu consigo apreciar um maravilhoso céu noturno repleto de estrelas espalhadas pelo universo.
**Sol:** Somos mesmo incríveis! Depois dessa merecemos aquele cafezinho.
- Terra e Sol saem do palco.

**Esquete 5: As Estações do Ano**

**Narrador:** No nosso planeta ocorrem quatro estações diferentes durante o ano: primavera, verão, outono e inverno. Contudo, na maioria das regiões do Brasil essas quatro estações não são bem definidas e se misturam em duas estações bem determinadas: uma quente e úmida e outra fria e mais seca. Mas o que causa as estações? E quanto ao dia claro e a noite, estes têm a mesma duração durante o ano?
- O Sol entra e se posiciona em um dos focos da eclíptica.

- Entram quatro personagens Terra, nomeadas como Terra 1, 2, 3 e 4, e se posicionam na eclíptica nos pontos de solstícios e equinócios, conforme apresentado nos livros didáticos. Cada uma das quatro Terras segura um cabo de vassoura em frente ao seu corpo de maneira inclinada em relação ao plano do palco (eclíptica) para representar a inclinação de 23,5 ° do eixo de rotação da Terra em relação a eclíptica.

**Sol:** Terra, uma coisa que eu venho observando enquanto você se move é que o seu eixo de rotação é inclinado em relação a eclíptica. Existe uma razão para isso?

**Terra 1:** Olha Sol, desde que eu me entendo por planeta Terra o meu eixo tem uma inclinação aproximada de 23,5º em relação a eclíptica. A causa exata disso eu não sei, mas essa inclinação provoca variações climáticas lindas em mim, chamadas de estações do ano, além de afetar a duração dos dias claros e das noites ao longo do ano.

**Sol:** Que máximo! Mas como isso é possível? Só por causa da inclinação do seu eixo de rotação?

**Terra 1:** Isso mesmo, e você tem um papel essencial nestes fenômenos. Vou explicar nos mínimos detalhes e prometo que você vai se encantar.

**Sol:** Vamos lá Terra, estou ansioso!

- Cada uma das personagens Terra fala sobre a sua posição, as datas correspondentes aos equinócios e solstícios, o início e o nome de cada estação.

- A Terra 1 fala sobre o equinócio de outono e aponta para baixo para indicar onde está o Hemisfério Sul.

**Terra 1:** Quando estou nesta posição ocorre o equinócio de outono e a data para isso no Hemisfério Sul é o dia 20 de março.

**Sol:** Para, para, para tudo Terra! Agora eu me embananei todo. O que é equinócio e o que é outono?

**Terra 1:** Calma Sol, vou explicar. O outono é uma das 4 estações do ano e o dia em que a mesma se inicia é chamado de equinócio, porque o dia claro e a noite possuem a mesma duração.

**Sol:** Mas por que isso acontece?

**Terra 1:** Porque os seus raios incidem sobre mim perpendicularmente ao meu eixo de rotação, de maneira que a quantidade de luz solar que eu recebo de você é a mesma em qualquer região.

**Sol:** Agora eu entendi! Mas o que caracteriza a estação do outono?

**Terra 1:** Nesta época do ano é comum observar as folhas das árvores caindo e muitas flores se transformarem em frutos. Com relação ao clima, a temperatura e a umidade do ar diminuem com o passar dos dias.

**Sol:** Que estação interessante! Mas ela ocorre em todo o planeta?

- A Terra 1 aponta para cima e para baixo para indicar os Hemisférios Norte e Sul, respectivamente.

**Terra 1:** Não! Como eu sou esférica, eu possuo dois hemisférios, Norte e Sul. Devido à inclinação do meu eixo de rotação, as estações são diferentes nas regiões dos dois hemisférios.

**Sol:** Entendi Terra, me diga mais!

- A Terra 2 fala sobre a estação do inverno:

**Terra 2:** Aproximadamente três meses depois, quando estou nesta posição, ocorre o solstício de inverno e a data para isso no Hemisfério Sul é o dia 21 de junho.

**Sol:** Espera aí Terra, assim você me confunde. Por que esse dia não é o equinócio de inverno? O que é solstício?

**Terra 2:** Na estação do inverno as noites tem maior duração que os dias claros. Como o primeiro dia do inverno é o dia do ano em que a noite é a mais longa de todas, nós damos o nome para ele de solstício.

**Sol:** Ah tá! Então os dias claros e as noites começam a ter durações diferentes após o equinócio de outono. Que interessante! E como é a estação do inverno?

**Terra 2:** Esta é a estação mais fria e seca do ano.

**Sol:** Mas por que essa estação é fria? Eu aqueço você com meus raios solares o ano todo!

**Terra 2:** Sim, mas por causa da inclinação do meu eixo de rotação, um dos meus hemisférios fica mais exposto aos seus raios do que o outro.

**Sol:** Mas é claro Terra! Isso significa que se o clima for mais frio no seu Hemisfério Sul, então o clima será mais quente no Hemisfério Norte. Mas qual é a estação com clima mais quente?

**Terra 2:** Calma que chegaremos lá!

- A Terra 3 fala sobre a estação da primavera:

**Terra 3:** Quando eu estou nesta posição ocorre o equinócio de primavera, também conhecido como equinócio vernal. A data para isso no Hemisfério Sul é o dia 22 de setembro.

**Sol:** Agora nem precisa explicar Terra, nesta data os dias claros e as noites também possuem a mesma duração, certo?

**Terra 3:** Isso mesmo Sol, pois é um equinócio.

**Sol:** E como é a estação da primavera?

**Terra 3:** Esta é a estação mais bela do ano, conhecida como estação das flores. As temperaturas são amenas, nem muito baixas e nem muito altas, e a umidade do ar aumenta, ficando um clima muito agradável.

**Sol:** Que lindo Terra! E quanto a estação mais quente do ano, você não vai falar sobre ela?

- A Terra 4 fala sobre a estação do verão:

**Terra 4:** Finalmente chegamos na estação que você queria Sol. Quando estou nessa posição ocorre o solstício de verão. A data de sua ocorrência no Hemisfério Sul é o dia 22 de dezembro, próximo do Natal.

**Sol:** Eu já entendi tudo Terra. Este dia é um solstício porque a duração do dia claro é diferente da noite. Estou certo?

**Terra 4:** Perfeitamente Sol.

**Sol:** E a estação do verão é mais quente porque no hemisfério em que ela ocorre, a exposição aos meus raios deve ser maior. Não é mesmo?

**Terra 4:** Sensacional Sol! Consequentemente, os dias claros são mais longos do que as noites.
**Sol:** Sim! Então, por causa da sua esfericidade, quando é inverno no Hemisfério Sul será verão no Hemisfério Norte e vice-versa.
**Terra 4:** Excelente observação Sol. Essa é uma das razões do porquê não tem sentido dizer que eu sou plana.
**Sol:** Credo! Mas, mesmo com todas as explicações fornecidas pela ciência sobre estes e outros fenômenos hoje em dia, tem gente que acredita nisso?
**Terra 4:** Infelizmente sim, em um movimento conhecido como terraplanismo. As pessoas terraplanistas são usualmente ingênuas e teimosas. Elas precisam estudar mais e enxergar a ciência como uma atividade investigativa séria e não como um conjunto de teorias conspiratórias.
- O Sol se vira para a audiência e diz:
**Sol:** É isso mesmo pessoal, todos nós precisamos estudar e entender o que é ciência. Ciência não é uma crença e nem uma religião, que você escolhe acreditar quando quiser ou conforme a sua conveniência. Ela é uma atividade profissional baseada em fatos, ou seja, em observações da natureza, e está em constante evolução.
- O Sol se volta para a Terra e diz:
**Sol:** Venha comigo Terra, vamos curtir nossa esfericidade e passear pela nossa galáxia, a maravilhosa Via Láctea.
- Sol e Terra saem do palco.
**Narrador:** Alterações climáticas vêm descaracterizando as estações do ano em diversas regiões do planeta como consequência de poluições, queimadas e desmatamentos. Além de catástrofes ambientais locais, a atividade humana tem contribuído significativamente para o aquecimento global. Investimentos em educação, ciência e tecnologia é o principal meio de conter este problema e prolongar a nossa sobrevivência no planeta.

**Esquete 6: A Lua e o seu Lado Oculto**

**Narrador:** A Lua é o maior e mais brilhante corpo celeste do nosso céu noturno e é essencial para a existência de vida na Terra. Muitas pessoas ainda não acreditam que o homem viajou à Lua, mas isso realmente aconteceu. Entre 1969 e 1972 com as missões Apolo, 12 astronautas norte-americanos caminharam sobre a superfície da Lua. Eles trouxeram para a Terra 382 kg de amostras de solo e rocha lunar para a realização de estudos científicos. Essas missões revolucionaram o nosso entendimento sobre a origem e outros mistérios do nosso satélite natural.
- A Terra entra, se posiciona no centro do palco e realiza o seu movimento de rotação.
- A Lua entra e gira no sentido horário em volta da Terra e em torno de si mesma. Quando a Lua completar sua órbita ela deverá ter realizado uma única revolução em torno de seu eixo, ou seja, esse movimento deve ser sincronizado. Para fornecer o efeito de manter a mesma face para a Terra, a personagem Lua precisa caminhar lateralmente em seu movimento de translação em torno da Terra. Para cada órbita completa da Lua, a personagem Terra deverá rotacionar 27 vezes. Marcações no palco podem ajudar na sincronização do movimento das duas personagens.

**Terra:** Oi Lua, como vai? Tem alguma novidade?

**Lua:** Nada de novo, continuo nesse ciclo sem fim, completando uma volta em torno de você a cada 27 dias terrestres em uma órbita elíptica.

**Terra:** Uau, isso significa que enquanto você completa a sua órbita eu giro em torno do meu eixo 27 vezes.

**Lua:** Sim, que tédio! Porque eu não consigo escapar por aí e dar uma volta pelo universo?

**Terra:** O Sol me explicou que os corpos celestes de maneira geral ficam próximos uns dos outros por causa da força gravitacional.

**Lua:** Por causa do que?

**Terra:** Por causa da força gravitacional. É uma força atrativa que exercemos uma sobre a outra de igual intensidade, mesma direção e sentidos contrários. Eu te puxo, você me puxa e nos mantemos próximas, é assim!

**Lua:** Mas se essas forças são de igual intensidade, por que sou eu que tenho que ficar orbitando você e não o contrário?

**Terra:** É por causa da nossa massa Lua. Os corpos celestes de menor massa orbitam os de maior massa, ou seja, os mais leves orbitam os mais pesados.

**Lua:** Isso até que faz sentido, porque quem orbita se movimenta mais e fica mais fitness!

**Terra:** Não é bem assim. Todos os corpos celestes estão em movimento em relação a um outro no universo.

**Terra:** Eu por exemplo, também tenho um movimento orbital em torno do Sol e ainda levo você comigo.

**Lua:** Então quer dizer que a massa do Sol é maior do que a sua?

**Terra:** Isso mesmo. Só para você ter uma ideia, a minha massa é 80 vezes maior do que a sua e a massa do Sol é em torno de 333 mil vezes maior do que a minha.

**Lua:** Minha nossa Terra, com uma massa dessa, qualquer planeta que pensar em passar "perto" do Sol irá orbitar ele.

**Terra:** Exatamente! E é por isso que chamamos o nosso sistema de sistema solar. Ele é composto por 8 planetas: Mercúrio, Vênus, Eu, Marte, Júpiter, Saturno, Urânio e Netuno; e outros 5 muito pequenos, chamados de planetas anões: Ceres, Plutão, Haumea, Makemake e Eris.

**Lua:** Que Sistema Maluco, mas ao mesmo tempo é fascinante!

**Terra:** Sim, ele é maravilhoso!

**Lua:** Terra, mas agora eu fiquei com uma dúvida. Se o Sol é tão mais pesado que você, porque eu orbito você e não ele, como os outros planetas? Eu poderia ser um planetinha anão muito bonitinho.

**Terra:** Mas você também orbita o Sol.

**Lua:** Claro que não, eu fico girando em torno de você e é por isso que eu não sou classificada como planeta, sou apenas o seu satélite natural.

**Terra:** Sim, mas como eu orbito o Sol e você me acompanha, você também está orbitando o Sol ué!

**Lua:** Certo, mas a minha dúvida é: por que eu não orbito o Sol como você? Mas eu já sei, nem precisa responder, deve ser por causa da distância. Por acaso a força gravitacional depende da distância?

**Terra:** Sim, quanto maior a distância menor é essa força.

**Lua:** Então é isso, pronto, resolvido. Eu orbito você porque a nossa distância é de 380 mil quilômetros, enquanto que o Sol está a uma distância de 400 vezes maior, a aproximadamente 150 milhões de quilômetros de nós.

**Terra:** Essa você errou Lua. Apesar da distância entre vocês ser bem maior do que a nossa, o Sol exerce em você mais do que o dobro da força gravitacional que eu exerço.

**Lua:** Mas gente, então por que raios eu não orbito o Sol como você?

**Terra:** Calma Lua! Isso ocorre por causa da sua velocidade.

**Lua:** Como assim?

**Terra:** Para você escapar da influência do meu campo gravitacional e viajar pelo espaço, você precisa de uma velocidade limite, chamada de velocidade de escape.

**Lua:** E quanto vale essa velocidade?

**Terra:** Um foguete, por exemplo, para ser lançado a partir da minha superfície precisa se mover a uma velocidade maior ou igual a 11,2 quilômetros a cada segundo (km/s).

**Lua:** Se ele se mover a uma velocidade menor do que essa ele não consegue chegar no espaço?

**Terra:** Isso mesmo!

**Lua:** Mas e no meu caso, eu não estou na sua superfície?

**Terra:** Quanto mais distante da minha superfície, menor é a velocidade de escape. No seu caso, você precisaria se mover a pelo menos 1,4 km/s para ficar livre de mim.

**Lua:** Puxa vida Terra, então é isso! Eu me movo a uma velocidade de 1,0 km/s. Afff... eu precisaria aumentar minha velocidade em 40%.

**Terra:** Sim. Com a sua velocidade você vai ficar me orbitando por um bom tempo, a não ser que algum outro corpo celeste te dê um baita empurrão.

**Lua:** Fazer o que né, se a nossa natureza é essa eu fico mais conformada de estar te acompanhando.

**Terra:** Sim, vamos juntas.

**Lua:** Terra, mas e o Sol, ele orbita alguém?

**Terra:** Sim, ele orbita um buraco negro que fica no centro da nossa galáxia, a Via Láctea.

**Lua:** É mesmo? Então esse cara deve ser supermassivo!

**Terra:** Sim Lua, a massa dele é milhões de vezes maior do que a massa do Sol e ele está muito, muito longe!

**Lua:** Ok, entendi.

- A Lua e a Terra se mantêm caladas por um tempo enquanto realizam os seus movimentos. A Terra nota que a Lua está sempre olhando para ela e acaba se incomodando.

**Terra:** Que bicho te mordeu Lua!? Por que você não para de me olhar?
**Lua:** Hein, o que? Ah, não é nada demais! Eu fico sempre com a mesma face voltada para você, porque o meu movimento de translação é sincronizado com o meu movimento de rotação.
**Terra:** Que interessante Lua. Então você tem uma face que eu não consigo enxergar?
**Lua:** Perfeitamente Terra!
**Terra:** Mas isso impede que existam dias claros e noites em você como em mim?
**Lua:** De maneira nenhuma! Assim como acontece com você, o Sol nasce e se põe no meu horizonte.
**Terra:** Mas e quanto a história de que você tem um lado escuro?
**Lua:** Eu não tenho um lado escuro, eu tenho um lado oculto, que você não consegue ver.
**Terra:** Mas isso não é a mesma coisa?
**Lua:** De jeito nenhum! O fato de você não ver o meu lado oculto não significa que ele seja escuro.
**Terra:** Então como é o seu dia Lua?
**Lua:** Ele é conhecido como dia lunar e tem duração de aproximadamente 709 horas, o que equivale a 29,5 dias terrestres.
**Terra:** Que dia longo, o meu tem duração de apenas 24 horas e eu achava que era muito.
**Lua:** Por isso que as vezes eu fico entediada.
**Terra:** Você sabia que existe um álbum de uma banda de rock progressivo chamada Pink Floyd com o nome "*The Dark Side of the Moon*", que significa "O lado escuro da Lua"?
**Lua:** Ah... então tá explicado. Esse negócio de lado escuro só podia ser invenção dos humanos. Eles são um bando de lunáticos e adoram romantizar e criar um ar de mistério para as coisas.
**Terra:** Enfim, mesmo você não tendo um lado escuro, podemos curtir as músicas desse álbum, porque elas são muito boas!
**Lua:** Com certeza! Solta o som...
- Lua e Terra saem do palco dançando e cantando parte da música *Brain Damage* da banda Pink Floid: "*I'll see you on the dark side of the Moon, ooh ooh ooh*".

**Esquete 7: As Fases da Lua e as Marés**

**Narrador:** Durante o dia lunar, cuja duração é de 29,5 dias terrestres, a aparência da Lua apresenta um ciclo regular de mudanças, caracterizado pelas fases nova, crescente, cheia e minguante. As fases da Lua estão diretamente relacionadas com as marés em nosso planeta. O conhecimento de sua periodicidade é muito importante para a realização de várias atividades costeiras, como surf, pesca, navegação e até mesmo para se preparar para tempestades.
- O Sol entra e se posiciona em um dos focos da eclíptica.
- A Terra entra e se posiciona no periélio.

- Entram quatro personagens Lua e se posicionam na órbita lunar em torno da Terra para representar as quatro fases: nova, crescente, cheia e minguante.
- A Lua Nova entra na frente das outras conversando com a audiência.

**Lua Nova:** E aí galera, agora eu vou mostrar para vocês que eu sou cheia de fases. Vejam só o meu brilho!

**Terra:** Que animação Lua. Existe alguma razão para isso?

**Lua Nova:** Claro minha amiga Terra, é porque eu estou na minha fase nova. Nesta fase eu começo a me revelar para você!

**Terra:** Sim, é como se você estivesse saindo aos poucos de traz da cortina escura do universo para eu te ver. Mas qual é a origem das suas fases Lua?

**Lua Nova:** É simples. Cada fase minha é determinada pela minha posição em relação a você e o Sol. E aí Sol como vai?

**Sol:** Oi Lua, oi Terra, tudo bem com vocês? Eu ouvi meu nome aí, sobre o que vocês estão conversando?

**Terra:** A Lua está explicando por que ela tem fases e parece que você tem um papel importante nisso.

**Sol:** Sério, como?

**Lua Nova:** Como eu estava dizendo, dependendo da minha posição em relação a vocês eu recebo uma quantidade de luz solar diferente e isso define minhas fases.

**Terra:** É mesmo Lua, assim como eu, você não tem brilho próprio, nós apenas refletimos a luz do Sol.

**Lua Nova:** Sim, se não fosse o Sol, nós viveríamos no escuro.

**Sol:** Mas fiquem tranquilas minhas amigas eu estou aqui para iluminar o caminho de vocês. Conte mais sobre suas fases Lua.

**Lua Nova:** Nesta posição que eu estou agora, entre o Sol e você, Terra, eu sou conhecida como Lua Nova.

**Terra:** Nesta fase eu só consigo te ver durante o dia Lua, você fica quase invisível no céu.

**Lua Nova:** Sim, mas eu vou surgindo aos poucos a cada noite e a parte que o Sol me ilumina forma uma letra 'C', quando sou vista do seu Hemisfério Sul, e uma letra 'D' quando sou vista do seu Hemisfério Norte.

**Terra:** Como assim Lua? Mas você continua com a mesma fase quando observada de ambos os hemisférios?

**Lua Nova:** Perfeitamente. Diferentemente das suas estações do ano, as minhas fases são as mesmas em ambos os hemisférios, a única diferença é que a minha aparência é invertida quando comparada nos dois hemisférios.

**Terra:** Isso faz sentido, porque se imaginarmos que o Hemisfério Norte está voltado para cima, o Hemisfério Sul estará voltado para baixo e um observador lá estará de cabeça para baixo (risos).

**Sol:** E o mesmo ocorre com o seu sentido de rotação e translação Terra. Se o Norte está para cima vocês orbitam em torno de mim no sentido anti-horário, mas se olharmos a partir do Sul, veremos o mesmo movimento no sentido horário.

**Lua Nova:** Risos..., que legal Sol. É por isso que vez ou outra nos movemos aqui no palco no sentido horário para explicar os fenômenos para a galera, pois estamos considerando um observador no Hemisfério Sul.

**Terra:** E isso é mais uma prova de que eu sou esférica e não plana (risos).

**Sol:** É isso aí! Mas continue falando de suas fases Lua, está muito interessante.

- A Lua Crescente continua:

**Lua Crescente:** Na fase crescente, minha face continua a ser revelada pela sua luz Sol. Durante a primeira semana, ou meu primeiro quarto de ciclo, metade da minha face pode ser vista. Ao final da segunda semana eu inicio a fase de Lua Cheia.

**Terra:** Áuuuuu... essa é a Lua do lobisomem (risos)!

**Sol:** Mas o que é lobisomem Terra?

- A Terra fala de maneira misteriosa:

**Terra:** É um homem que se transforma em lobo em noites de Lua Cheia. Ele sai em busca de vítimas para poder alimentar-se do sangue delas.

**Sol:** Parece assustador!

- A Lua Cheia continua sua fala demonstrando irritação.

**Lua Cheia:** Sim, mas eu não tenho nada com isso!

**Terra:** Calma Lua, o lobisomem é só uma criatura folclórica inventada pelos humanos.

**Lua Cheia:** Esses humanos! Sempre fazendo gracinhas para o meu lado. Bando de lunáticos!

**Terra:** Mas e depois Lua? Quando você está na fase Cheia o seu ciclo recomeça?

**Lua Cheia:** Não, minha fase Cheia indica que o meu ciclo ainda está na metade.

- A Lua Minguante continua:

**Lua Minguante:** No terceiro quarto do meu ciclo inicia a minha fase Minguante e após duas semanas eu começo o ciclo novamente com a fase Nova, que ocorre quando eu e o Sol estamos praticamente na mesma posição no céu, quando vistos por você Terra.

**Terra:** Pelas minhas contas Lua o seu ciclo demora 4 semanas, ou seja, um mês. É isso mesmo?

**Lua Minguante:** Na verdade são 29,5 dias.

**Terra:** Mas o tempo para você completar sua órbita não é em torno de 27 dias? Agora eu fiquei confusa!

**Lua Minguante:** Sim, este é o tempo para eu completar uma revolução completa em torno de você e é conhecido como mês sideral. O tempo para eu completar o meu ciclo de fases é dois dias mais longo e é conhecido como mês sinódico ou lunação.

**Terra:** Agora que você falou sideral eu entendi tudo. O mês sinódico é maior do que o sideral pela mesma razão que o meu dia solar é maior que o meu dia sideral.

**Lua Minguante:** Eu não conseguiria explicar melhor Terra.

**Sol:** E você Terra, também possui fases?

- A Terra dá ênfase em sua fala.
**Terra:** Eu não diria faaaaaases, como a Lua tem, mas eu sinto o nível das águas dos meus oceanos mudarem à medida que a Lua realiza o ciclo de fases dela.
- A Lua Crescente retoma a fala.
**Lua Crescente:** Isso significa que eu exerço alguma influência em você durante o meu movimento de translação?
**Terra:** Sim, e isso ocorre diariamente!
**Sol:** Explica para gente como é isso Terra.
**Terra:** Aproximadamente três quartos da minha superfície é coberta por água e 98% dessa água está nos oceanos, formando a minha hidrosfera.
**Sol:** De toda a sua água apenas 2% é água doce?
**Terra:** Isso mesmo Sol.
**Lua Crescente:** Então se esta água for poluída, a vida de muitas espécies em você será comprometida Terra, porque não dá para beber água salgada, certo?
**Terra:** Bem observado Lua e é por isso que os meus lagos e rios precisam ser protegidos e bem cuidados.
- O Sol se manifesta interagindo com a audiência.
**Sol:** Muito bom Terra, acredito que todos aqui sabem que isso é muito importante. CERTO PESSOAL? Mas e quanto a influência da Lua nas suas fases ou sei lá o que?
**Terra:** A flutuação diária no nível dos meus oceanos são chamadas de marés. Existem duas marés baixas e duas marés altas por dia na maioria de minhas regiões costeiras, que podem variar de alguns centímetros a muitos metros.
**Sol:** Ok Terra, eu entendi o que são as marés, mas o que as causam?
- A Lua chama a atenção do Sol para o que a Terra irá dizer:
**Lua Crescente:** Eu aposto com você que ela virá com a história de força gravitacional novamente.
**Terra:** É isso mesmo Lua.
**Lua Crescente:** Ahaaa, não falei para você. É sempre a mesma história.
**Sol:** Mas agora eu fiquei confuso Terra. Se o lance é a força gravitacional, por que as suas marés estão relacionadas com as fases da Lua e não comigo?
**Lua Crescente:** Boa Sol!
**Sol:** Digo isso porque a força gravitacional que a Lua exerce em qualquer ponto da sua superfície é cerca de cem vezes menor do que a força gravitacional que eu exerço.
**Terra:** Você fez a lição de casa Sol, isso mesmo. Mas as marés são causadas principalmente pelas variações das forças gravitacionais que vocês dois exercem sobre os meus oceanos.
**Sol:** Então a minha influência continua sendo maior, obviamente.
**Terra:** Neste caso não, pois como você está muito mais longe de mim do que a Lua, as variações do seu campo gravitacional são menos perceptíveis, enquanto que a não homogeneidade do campo gravitacional da Lua em toda a minha superfície é consideravelmente maior.

**Lua Crescente:** Viu só Sol, para provocar as marés na Terra tem que ter jogo de cintura (risos).
- O Sol se mostra indignado com a situação:
**Sol:** Mas a Terra disse que eu também influencio ué!
**Terra:** Sim, mas as marés induzidas pela Lua são aproximadamente o dobro das marés induzidas por você.
**Sol:** Então tá, mas qual a relação da altura das suas marés com as fases da Lua?
**Terra:** Quando nós três estamos alinhados, na Lua Nova e na Lua Cheia, os efeitos gravitacionais são reforçados e eu apresento as maiores marés possíveis.
- A Lua Nova ou a Cheia podem assumir a fala.
**Lua Cheia:** E eu aposto que tem um nome diferente e espalhafatoso para isso.
**Terra:** Estas são conhecidas como marés de Sizígia ou de águas-vivas.
**Sol:** Tem razão Lua, quanto nome para a mesma coisa!
**Terra:** E quando a Lua está finalizando o primeiro e o terceiro quartos do seu ciclo, com as fases Crescente e Minguante, minhas marés são menores e são conhecidas como marés de quadratura ou de águas-mortas.
**Sol:** Fascinante! A cada dia que passa eu fico mais encantado com a relação natural que existe entre nós.
**Terra:** E vocês sabiam que as marés também ocorrem na parte da minha superfície, onde há terra, e na minha atmosfera?
**Lua Cheia:** Eu até acredito que existam, mas isso deve ser muito difícil de observar, certo?
**Terra:** Sim, as marés terrestres e atmosféricas só podem ser detectadas por instrumentos científicos muito sensíveis.
**Sol:** A ciência é realmente incrível!
- Saem do palco o Sol, a Terra e as Luas.

**Esquete 8: Eclipses Lunares e Solares**

**Narrador:** De tempos em tempos é anunciado em diversos canais de notícias que um eclipse irá acontecer e que o mesmo poderá ser melhor apreciado em determinadas regiões do planeta. Mas o que são e o que causam os eclipses? Qual a frequência que os mesmos ocorrem? Que tipos de eclipses existem?
- O Sol entra e se posiciona em um dos focos da eclíptica.
- A Terra entra e se posiciona no periélio.
- A Lua entra e varia suas posições na órbita lunar conforme a necessidade. Esta órbita é representada por um colar de isopor utilizado pela Terra.
- Nesta cena os corpos celestes são representados principalmente pela cabeça dos estudantes-atores.
**Lua:** Uaaaau Terra que colar bonito!
**Terra:** Sim, estou utilizando ele para representar o plano de sua órbita em torno de mim.

**Lua:** Que ótimo, dessa forma fica mais fácil para a galera entender o que vamos discutir nesta cena.

**Terra**: Lua, você nos explicou que as suas fases são determinadas pela quantidade de luz solar refletida na sua superfície, certo?

**Lua:** Perfeitamente Terra e isso ocorre por causa da minha posição no céu em relação a você e o Sol.

**Terra:** Mas eu ainda não entendi o que provoca o seu lado escuro. Isso é um efeito da minha sombra sobre você?

**Lua:** Não, isso é causado pela minha própria sombra, ou seja, é a minha face oposta ao Sol.

**Terra:** Entendi, então é semelhante à forma que as minhas noites ocorrem.

**Lua:** Isso mesmo, isso acontece com qualquer objeto, seja este celeste ou não. A sua sombra sobre a minha superfície tem outro nome.

**Terra:** Qual?

**Lua:** É um fenômeno chamado de Eclipse Lunar.

- O Sol interrompe a conversa dizendo:

**Sol:** Opa! Falou em eclipse é comigo mesmo.

**Terra:** Você sabe o que é eclipse Sol?

**Sol:** Sim, é uma mania que alguns corpos celestes têm de ficar bloqueando a minha luz para fazer sombra nos outros.

**Terra:** Ah tá, eu já vi a sombra da Lua passando pela minha superfície algumas vezes. Então o fenômeno que ocorre aqui é o Eclipse Terrestre?

**Sol:** Não Terra, este é o Eclipse Solar.

**Terra:** Como assim, qual a diferença entre os dois?

**Sol:** Você permite que eu explique Lua?

**Lua:** Fique à vontade Sol. Afinal de contas, o brilho que é bloqueado em um eclipse é o seu (risos).

- O Sol explica de maneira pomposa o que é eclipse para a Terra.

**Sol:** Muito bem! Os eclipses que ocorrem entre nós três é um fenômeno natural causado pelo nosso alinhamento no espaço. Quando a Lua está entre nós ocorre o eclipse solar e quando você está entre mim e a Lua ocorre o eclipse lunar.

- A Terra contesta a explicação do Sol.

**Terra:** Se esta é a causa, então deveria ocorrer eclipses entre nós pelo menos duas vezes por mês, porque ficamos alinhados na Lua Nova e na Lua Cheia e eu não vejo isso. Você concorda Lua?

- A Lua fala de maneira tímida:

**Lua:** Sabe o que é Terra, eu não te falei antes, mas na verdade nosso alinhamento nas minhas fases Nova e Cheia não é, na maioria das vezes, aquela coisa bonita, certinha, ele é meio aproximado.

**Terra:** Como não Lua! Todos os meses você tem as fases Nova e Cheia ué.

**Sol:** Isso mesmo, agora eu é que fiquei confuso!

**Lua:** Sim, mas independentemente disso, nosso alinhamento não é perfeito. Se traçarmos uma linha entre nós três, alguém vai ficar de fora.
- A Terra se mostra impaciente.
**Terra:** Mas como isso é possível? Se você passa entre mim e o Sol e depois eu fico entre você e o Sol estamos alinhados.
- O Sol também se manifesta.
**Sol:** Claro, eu também vejo isso!
- A Lua utiliza o colar de isopor da Terra para mostrar o efeito da inclinação de sua órbita.
**Lua:** Vou explicar utilizando o seu colar Terra, que representa o plano da minha órbita em torno de você. A dificuldade de nos alinharmos perfeitamente ocorre porque a minha órbita possui uma inclinação de aproximadamente 5 ° em relação à eclíptica.
**Terra:** Isso até que faz sentido, porque às vezes eu tenho a impressão de que você está um pouco acima ou abaixo do Sol quando você passa entre nós.
- A Lua muda sua posição, abaixando-se e levantando-se, para mostrar porque não ocorrem eclipses todas as vezes que os três corpos celestes estão na mesma direção. A Terra toma o cuidado para sempre manter o colar alinhado com a cabeça da personagem Lua em todas as posições que ela ocupa. Para um melhor efeito, é interessante que o estudante que faz a personagem Lua seja mais alto que a personagem Sol.
**Lua:** Sim, vejam só: quando eu estou acima do Sol minha sombra passa por cima de você e quando estou abaixo, minha sombra passa por baixo e não ocorre nenhum eclipse.
**Sol:** Então todas as vezes que você cruzar o plano da eclíptica, para cima ou para baixo, haverá um eclipse lunar ou solar, certo?
**Terra:** É verdade Lua, porque quando isso acontecer nosso alinhamento será perfeito.
**Lua:** Essa é a ideia. Contudo, existe mais um detalhe que é necessário levar em conta.
- O Sol sussurra para a Terra dizendo:
**Sol:** Eu estou começando a achar que a Lua é muito complicadinha viu.
**Terra:** Eu também Sol. Que detalhe é esse Lua?
**Lua:** Eu ouvi viu Sol. Não há complicação nenhuma é só a minha natureza. Além de ser inclinada, a minha órbita também varia a direção da inclinação em relação a você, à medida que a Terra e eu te orbitamos.
- A Terra movimenta o disco de isopor em torno do pescoço como se fosse um bambolê dizendo:
**Terra:** Ah Lua, então a sua órbita fica mudando a direção em torno de mim como se fosse um bambolê?
**Lua:** Fabuloso Terra, adorei a analogia.
**Terra:** Que legal! E como ocorrem exatamente os eclipses lunares e solares? Com todos esses detalhes eu ainda não entendi.
**Lua:** Você quer continuar sua explicação Sol?
**Sol:** Diante das novidades em relação à inclinação da sua órbita e o bamboleio da mesma, é melhor você continuar.

**Lua:** Certo. Os eclipses lunares ocorrem na minha fase Cheia quando você, Terra, está precisamente entre mim e o Sol, ou seja, quando o nosso alinhamento é perfeito.

- O Sol complementa a explicação da Lua.

**Sol:** E digo mais, quando isso ocorre a Lua fica completamente escondida na sua sombra, a ponto de eu não conseguir vê-la.

**Terra:** Sim, mas os seus raios solares espalham-se em mim e continuam iluminando a Lua, deixando ela com uma cor avermelhada.

**Sol:** Que interessante, é uma pena que eu não consiga ver você dessa forma Lua. Mas por que isso acontece?

**Lua:** Essa para mim é nova. Conta para gente Terra, por que isso acontece?

**Terra:** Isso é um efeito da minha atmosfera. Quando a luz do Sol passa por ela a cor azul se espalha mais facilmente e é por isso que o meu céu fica azul durante o dia. As cores vermelha e laranja são mais difíceis de serem espalhadas e atravessam a minha atmosfera até chegar na Lua, dando a impressão de que ela está avermelhada quando eu a observo.

**Sol:** Que magnífico Terra, você é um planeta cheio de maravilhas.

**Terra:** Obrigado Sol! Esse efeito também faz com que o meu pôr do Sol seja mais alaranjado.

**Lua:** Nossa Terra, quando você falou cor avermelhada eu já pensei que alguma coisa poderia estar errada comigo, uma alergia ou algo do tipo.

**Sol:** Não Lua, pelo que eu entendi você não muda de cor. O tom avermelhado é apenas uma cor aparente observada pela Terra.

**Terra:** Exatamente! Mas este receio não é só coisa sua, Lua. Antigamente os humanos acreditavam que a lua vermelha era um aviso de guerra, maus presságios ou que uma catástrofe natural iminente iria acontecer.

**Lua:** Pobres humanos, provavelmente eles já estariam extintos se não fosse o avanço da ciência.

**Sol:** Com certeza! Ainda mais agora com esse tanto de doença que tem por aí, existe até pandemia. Imagina se não existissem as vacinas!

- A Terra fala com tristeza:

**Terra:** Sim, mas as vezes eu penso que eu seria bem mais saudável e bonita sem eles.

**Sol:** Calma Terra, não fique triste! Com o tempo eles vão se familiarizar e confiar mais na ciência e tudo vai melhorar.

**Terra:** Espero que isso ocorra o quanto antes, porque a aparência vermelha da Lua está surgindo mesmo quando não há o eclipse lunar. Às vezes ela fica até meio amarelada.

**Sol:** Mas por que isto está acontecendo Terra?

**Terra:** Por causa da poluição da minha atmosfera, mas isso só pode ser notado quando a Lua está próxima do meu horizonte.

**Lua:** Nossa Terra, então as minhas cores aparentes estão se tornando um sinal realmente ruim.

**Sol:** É verdade, mas precisamos continuar acreditando que a ciência irá mudar o pensamento das pessoas para melhor.

**Terra:** Sim. Enfim, conte-nos mais sobre os eclipses Lua.
**Lua:** Bom, quando estou na fase Cheia e minha cor fica avermelhada, quando vista da Terra, significa que o eclipse lunar é total.
**Sol:** Mas por que total? Existem outros tipos de eclipse lunar?
**Lua:** Sim, os eclipses também ocorrem de maneira parcial, quando apenas uma parte da sombra da Terra passa pela minha superfície.
- O Sol concorda bamboleando sua mão.
**Sol:** É verdade, por causa da inclinação e do bamboleio da sua órbita nem sempre nosso alinhamento é perfeito.
**Terra:** Eu até entendo isso Sol, mas como dá para saber que isso não é a sua própria sombra Lua em uma fase Crescente ou Minguante?
**Lua:** Por que é possível notar a sombra na minha superfície crescer e diminuir em pouco tempo. Apesar de sua sombra ser grande, por causa do seu tamanho em relação ao meu, eu consigo sair totalmente dela em apenas algumas horas.
**Terra:** É verdade, você está sempre em movimento.
**Sol:** Mas e quanto ao eclipse solar, Lua, ele também ocorre de maneira total e parcial?
**Terra:** Eu acho que não Sol, porque eu sempre te vejo.
**Lua:** O eclipse solar é resultado do nosso alinhamento perfeito quando estou na fase Nova, entre o Sol e você Terra. Ele é total em apenas algumas regiões da sua superfície.
**Terra:** Mas por quê?
**Lua:** Por que eu sou bem menor do que você, minha sombra não é suficiente para cobrir toda a sua superfície, como ocorre no eclipse lunar total.
**Terra:** E você saberia me dizer o tamanho da sua sombra na minha superfície?
**Lua:** Ela mede em torno de 480 km de largura.
**Sol:** E quanto as pessoas que estão nessa região de sombra, o que elas veem?
**Lua:** A sombra que se forma na Terra consiste de duas partes: a parte central e mais escura, chamada de umbra, e a parte mais externa e menos escura, chamada de penumbra.
**Sol:** Por que essa classificação? Na região de sombra não é tudo escuro?
**Terra:** É verdade Lua, para mim sombra é tudo a mesma coisa.
**Lua:** Não! A luz do Sol não consegue penetrar na umbra e por isso, as pessoas que estão nessa região veem o dia claro se tornar noite.
**Terra:** Isso significa então que as pessoas na umbra vão ver um eclipse solar total, certo?
- O Sol complementa fazendo pausas na sua fala demonstrando dúvida.
**Sol:** E as pessoas que estão na penumbra irão ver um eclipse solar parcial?
**Lua:** Exatamente, vocês entenderam tudo!
**Sol:** E quanto dura um eclipse solar?
**Terra:** Eu vejo a sombra da Lua na minha superfície por apenas alguns minutos Sol.
**Lua:** Apesar disso, a Terra rotaciona tão rápido que a minha sombra varre um terço do caminho ao redor da superfície dela, antes que o alinhamento entre nós três acabe.
**Sol:** Que interessante Lua. Você saberia dizer qual a frequência de ocorrência dos eclipses lunares e solares?

**Terra:** Eu também tenho essa dúvida, mas agora sabemos que o eclipse solar não ocorre todas as vezes que a Lua está na fase Nova e o eclipse lunar não ocorre em todas as Luas Cheias.

**Lua:** Muito bom Terra. Os eclipses lunares, de qualquer tipo, podem ser observados de 2 a 3 vezes por ano.

**Sol:** Sensacional Lua, isso é mais de um eclipse por ano.

**Lua:** Já os eclipses solares podem ser observados de 2 a 5 vezes por ano em algum lugar da Terra.

**Terra:** Mas esses números variam muito. Existe alguma periodicidade para a ocorrência dos eclipses, como o mês sinódico, que é o período em que as suas fases se repetem?

**Lua:** Sim, este é chamado de Ciclo de Saros. A cada 6.585 dias, ou 18 anos, aproximadamente, a posição de nós três no espaço volta a ser a mesma e todos os eclipses lunares e solares observados nesse período serão observados novamente.

**Terra:** Então é por isso que os humanos conseguem prever os eclipses, os cientistas conhecem esse ciclo.

- O Sol fala de maneira descontraída:

**Sol:** Isso é muito incrível. Mas imaginem só o desespero dos humanos ao verem o dia claro se tornar noite em questão de minutos (risos).

**Lua:** (Risos), sim, eles devem achar que é o fim do mundo.

**Terra**: Antes do surgimento da ciência moderna isso era muito comum, mas com o avanço da ciência e o desenvolvimento de novas tecnologias, os cientistas têm mostrado e explicado fenômenos incríveis da natureza.

**Sol:** Terra, se a ciência é algo tão poderoso e esclarecedor, por que os humanos insistem em menosprezar o que os cientistas dizem?

- A Lua responde em um tom jocoso:

**Lua:** É porque esse povo é cabeça dura, Sol!

**Terra:** Não é bem assim Lua. Algumas pessoas são teimosas mesmo, mas a verdade é que existe muita desigualdade entre os humanos, de maneira que alguns têm exercido o domínio sobre os outros através da manipulação de informação.

**Sol:** Mas como isso pode ser revertido Terra?

- A Lua continua com um tom de brincadeira.

**Lua:** É só eliminar todos os manipuladores e mentirosos que utilizam a informação conforme a conveniência.

- Todos riem.

**Terra:** Não é necessário nenhum tipo de violência Lua.

**Lua:** O que pode ser feito então?

**Terra:** Basta investir em uma educação igualitária, laica e pública de qualidade.

**Lua:** Concordo com você Terra e para isso é necessário que as pessoas valorizem as escolas e universidades públicas, porque estas instituições fornecem um dos principais meios para mediar a convivência entre as classes sociais.

**Sol:** Eu estou convencido de tudo isso, o meu voto é de vocês, e digo mais: No estado atual que estamos vivendo, a ciência é o principal empreendimento para que os seres humanos consigam conviver em harmonia com a natureza.

**Terra:** Isso mesmo Sol. Além dos investimentos em educação, precisamos confiar no avanço da ciência para continuarmos apreciando e desvendando os mistérios do nosso SISTEMA MALUCO!

- Saem do palco a Lua, a Terra e o Sol e em seguida entram todos e agradecem ao público.

Fim do espetáculo!